\documentclass[11pt, a4paper]{article} %

\usepackage{graphicx} %
\usepackage[utf8]{inputenc} %
\usepackage{textcomp}
\usepackage{xcolor} %
\usepackage{listings} %
\usepackage{amsfonts} %
\usepackage{amsmath} %
\usepackage{amssymb} %
\usepackage{mathrsfs} %
\usepackage{latexsym} %
\usepackage{euscript} %
\usepackage{cite} %
\usepackage{mciteplus}

\usepackage{hyperref}
\definecolor{esablue}{HTML}{0081B8}
\hypersetup{
	colorlinks,%
	citecolor=esablue,%
	filecolor=esablue,%
	linkcolor=esablue,%
	urlcolor=esablue,
	bookmarks=true,
	bookmarksopen=true,
	pdftitle=A simple and effective parametrisation for Earth-impacting orbits,
	pdfauthor={Alexandre Payez},
	pdfdisplaydoctitle
}

\usepackage{array}

\usepackage[normalem]{ulem}

\usepackage{geometry} 
\usepackage[font=small]{caption}

\usepackage[small]{titlesec}

\makeatletter
\def\@maketitle{%
  \newpage
  \null
	{\centering
  \let \footnote \thanks
    {\Large \textsc \@title \par}%
    \vskip .8em%
    {\normalsize
      \lineskip .5em%
      \begin{tabular}[t]{c}%
	\small
        \@author
      \end{tabular}\par}%
    \vskip 1em%

	}
  \par
  \vskip 0em}
\makeatother

\newcommand{\be}{\begin{equation}}
\newcommand{\ee}{\end{equation}}
\newcommand{\atan}{\mathrm{atan}}
\newcommand{\acos}{\mathrm{acos}}

\usepackage{caption}
\title{%
		A simple and effective parametrisation\\for Earth-impacting orbits
}

\author{%
	Alexandre~Payez\thanks{\href{mailto:alexandre.payez@esa.int}{alexandre.payez@esa.int}, \href{mailto:alexandre.payez@desy.de}{alexandre.payez@desy.de}}
	\\[.2cm]
	\small{\emph{European Space Agency, ESOC, Robert-Bosch-Str.\@ 5, D-64293 Darmstadt, Germany}}
}

\begin{document}

\maketitle

\begin{abstract}

This work presents a new parametrisation suitable for parameter-space studies of heliocentric Earth-impacting orbits.
Originally motivated by
the issue of potentially hazardous asteroids (PHAs) and the mitigation of such a risk, we show that the simultaneous analysis of all the conceivable impacting elliptical orbits is greatly facilitated by the use of a parametrisation that involves only the true anomaly at impact, the eccentricity and the inclination.

While the new parametrisation is presented from an explicit planetary-defence perspective, it is general enough to be useful in analogous studies.

\end{abstract}

\section{Introduction}

Collisions take place in our Solar System.
The craters observed on the surface of moons and planets, due to impacts with minor celestial bodies such as comets and asteroids, act as so many reminders.
The most spectacular outcomes of past collisions with our planet not only include the extinction of the dinosaurs at the end of the Cretaceous~\cite{Alvarez_etal:1980,SmitHertogen:1980,Schulte_etal:2010},
but probably also the formation of the Moon~\cite{Asphaug:2014};
it is moreover conceivable that material brought by cometary impacts played a role in the very development of life on Earth, as discussed {\it e.g.\@} in Refs.~\cite{Oro_etal:1991,Jenniskens_etal:2004,Altwegg_etal:2016} and references therein.

Whereas the Earth is continuously bombarded by meteoroids entering the atmosphere as mostly harmless shooting stars~\cite{Ceplecha_etal:1998,Nesvorny_etal:2010},
collisions with the largest asteroids and comets (bigger than 1~km) are exceedingly rare\hspace{1pt}---\hspace{1pt}involving timescales longer than the existence of our own species~\cite{Oepik:1976,HarrisDAbramo:2015}. 
Since large objects are moreover easier to detect, they are in fact even more unlikely to catch us off-guard, seeing that \mbox{near-Earth} objects (NEOs)
are being continuously searched for and actively monitored. Large parts of the sky are indeed surveyed, in visible and infrared wavelengths, as well as with radars~\cite{SurveysReview:2015};
the data are centralised in the IAU Minor Planet Center (MPC) database~\cite{MPC} and made freely available to all.
The impact threat assessment~\cite{TorinoScale,PalermoScale} for each newly discovered comet or asteroid is then notably done independently by the NASA-JPL Sentry system and the ESA-sponsored NEODyS service, and is constantly reevaluated as more observations become available.

There is currently an ongoing international effort to find most of the so-called potentially hazardous asteroids (PHAs): asteroids larger than $\sim 140$~m with orbits closely approaching that of the Earth, by less than 0.05~astronomical units (AU). 
Such an object would indeed represent an undeniable threat should it be found on a collision course with our planet.
Over the last two decades, the number of discovered PHAs increased from about a hundred to almost two thousands, 
with no such impact foreseen to happen within the next 100 years.
Since more than 90\% of the very largest near-Earth asteroids are already known~\cite{Mainzer_etal:2011},
the most plausible threat comes from currently unknown asteroids of dimensions of a few tens (much more likely)
to several hundreds of metres~\cite{HarrisDAbramo:2015}.
If such a rare\footnote{The estimated average impact interval for 140-m objects is larger than 10,000 years~\cite{HarrisDAbramo:2015}.} event were to happen, it
could have consequences correspondingly ranging from local to large-scale destruction~\cite{DefendingPlanetEarth:2010}.
In fact, significant local damage can be dealt even when the meteor disintegrates in the atmosphere before ever reaching the ground:
often taken as an
example was the Tunguska event in Siberia on June 30th 1908~\cite{Farinella_etal:2001}; 
a more recent instance being the airburst over Chelyabinsk in 2013, caused by the atmospheric entry of an asteroid of about ten thousand tons (diameter of about 20 m)~\cite{Borovicka_etal:2013}.

A natural question is whether a collision could be avoided if it is identified early enough.
Besides the obviously needed civil response to such an extraordinary event,
a number of space missions could indeed be envisaged\hspace{1pt}---\hspace{1pt}not only fly-by and rendez-vous missions to both reduce the orbital uncertainties and characterise the object {\it in situ}, but mitigation missions as well. 
Proposed mitigation concepts typically aim at imparting enough momentum to avoid an impact ({\it e.g.\@} with kinetic impactors or gravity tractors)
but might also involve the complete annihilation of the impacting body~\cite{DefendingPlanetEarth:2010}.
As there can be significant uncertainties on the momentum and energy transfers,
conducting a first mitigation demonstration on a safe object is in fact a prime goal endorsed by the Space Mission Planning Advisory Group (SMPAG).\footnote{SMPAG has a mandate from the United Nations to issue recommendations on NEO threat mitigation whereas the International Asteroid Warning Network (IAWN) is trusted with the NEO observations~\cite{IAWNSMPAG:2013}.}  
Such a mission, like AIDA~\cite{AIDA:2015} or NEOT$\omega$IST~\cite{NEOTWIST:2016},
would moreover find itself in the continuity of
a large number of varied missions to asteroids and comets\hspace{1pt}---\hspace{1pt}among which 
OSIRIS-REx~\cite{OSIRIS-REx:2014}, the two Hayabusa missions~\cite{Hayabusa:2006,Hayabusa2:2013}, Deep Impact~\cite{DeepImpact:2005},
and Rosetta~\cite{Rosetta:2007}.

From the mission-analysis viewpoint, the issue of potentially hazardous objects (PHOs) therefore provides a strong incentive to solve 
many kinds of astrodynamical problems and to devise trajectories to hypothetical Earth-impacting asteroids or comets.
Interestingly, rather than studying only one or two threats individually, these problems have sometimes been approached at the orbital parameter-space level,
most notably in Refs.~\cite{SanchezMcInnes:2012, FB:2015, ThiryVasile:2016}
and also to some extent in Refs.~\cite{ParkRoss:1999, RossParkPorter:2001, ParkMazanek:2005, VasileColombo:2008, CasalinoSimeoni:2012, Carusi_etal:2002}.
Based on samples of fictitious crossing orbits, 
such works can reveal how different mission properties depend on the PHO orbital elements.
One should of course reckon that these more general studies typically consider a few reasonable simplifications, 
{\it e.g.\@}
closed keplerian orbits around the Sun~\cite{ParkRoss:1999, RossParkPorter:2001, ParkMazanek:2005, VasileColombo:2008, SanchezMcInnes:2012, CasalinoSimeoni:2012,FB:2015,ThiryVasile:2016},
strict impacts~\cite{ParkRoss:1999, RossParkPorter:2001, Carusi_etal:2002, ParkMazanek:2005, VasileColombo:2008, SanchezMcInnes:2012, CasalinoSimeoni:2012,FB:2015,ThiryVasile:2016},
and a circular Earth orbit~\cite{ParkRoss:1999, RossParkPorter:2001, ParkMazanek:2005, SanchezMcInnes:2012, CasalinoSimeoni:2012, FB:2015, ThiryVasile:2016}.
While the results will still hold in more realistic settings, the aim is not to be precise enough for an actual mission.
In return, going beyond isolated cases is what enables a better understanding of the overall problem\hspace{1pt}---\hspace{1pt}a great benefit when most near-Earth objects are yet to be discovered~\cite{HarrisDAbramo:2015}.
This kind of approach is in fact what directly motivated the current work. 

In this article, we present a 
simple
and
natural
parametrisation 
precisely
designed 
for para\-meter-space studies of heliocentric Earth-crossing orbits with strict impacts.
To make its usefulness in a practical context as clear as possible, the presentation will be done assuming a planetary-defence application, even though this parametrisation turns out to be general enough to prove useful in a wider range of problems.
The paper is organised as follows.
Section~\ref{sec:param}
first
discusses
why an alternative
is
of interest for this kind of studies.
The new parametrisation is then introduced in Sec.~\ref{sec:newparam},
and is shown to greatly facilitate
the simultaneous analysis of all the conceivable
impacting elliptic orbits.
The advantages of a polar-plot representation of the new parametrisation are then the subject of Sec.~\ref{sec:polar}.
Finally, further practical uses of interest for planetary defence are discussed in Sec.~\ref{sec:furtheruses}, before we conclude.

\subsection*{Assumptions and scope}

	No attempt to distinguish between asteroids and comets is made in this work; different systems of classification exist, see {\it e.g.\@} Ref.~\cite{Jewitt:2012}\hspace{1pt}---\hspace{1pt}and there arguably is a continuum~\cite{Hsieh:2017}.
	Focusing solely on the orbits themselves, 
	any minor celestial body on a collision course with our planet will then be referred to as ``PHA'' or ``asteroid'' for simplicity.

For clarity,
the assumptions made throughout this paper are that:
\begin{enumerate}
	\item PHA orbits can be described by elliptic orbits around the Sun (2-body problem);\footnote{Limited to bounded orbits here, the parametrisation can be extended to parabolic or hyperbolic cases.}
	\item the minimum orbit intersection distances (MOIDs) are small enough to be neglected;
	\item the Earth orbit is essentially circular.
\end{enumerate}
	In the absence of prior close-encounters~\cite{Carusi_etal:2008}, 
	these are good approximations for describing the heliocentric motion of fictitious 
	asteroids or comets
	foreseen to impact at most a few decades later; 
	they are often discussed in the literature (individually or all at once), as in {\it e.g.\@} Refs~\cite{ParkRoss:1999, RossParkPorter:2001, Carusi_etal:2002, ParkMazanek:2005, VasileColombo:2008, SanchezMcInnes:2012, CasalinoSimeoni:2012,FB:2015,ThiryVasile:2016, Bonanno:2000, Izzo:2007, SanchezMcInnes:2011, BombardelliBau:2012}.

\section{The usual way of parametrising orbits strictly impacting Earth}\label{sec:param}

Any
impending 
collision of a PHA with our planet must necessarily happen 
as it crosses the orbital plane of the Earth\hspace{1pt}---\hspace{1pt}that is, either around its ascending or its descending node if the orbit is inclined.
An inertial frame built upon the ecliptic plane then clearly represents a judicious choice for studying the parameter space of 
generic crossing orbits.

If the Earth orbit is moreover considered to be circular for simplicity,
one 
gains the freedom to choose 
where on the circle
the impact happens.
Due to the increased symmetry,
the physics
then
indeed becomes completely independent of
such a choice:
all the possible 
crossing
locations
are then
indistinguishable from one another.
For instance~\cite{FB:2015}, the
impact can always be 
chosen 
to happen along the $x$-axis without any loss of generality:\footnote{Focusing explicitly {\it e.g.\@} on the next impact location removes the ambiguity for double-crossing orbits.}
\be
	\vec{r}_I = 1~{\rm AU} \ \vec{e}_x;
	\label{eq:impactlocation}
\ee
one can then complete the orthonormal basis by setting $\vec{e}_y$ along the Earth velocity at that point, and $\vec{e}_z$ parallel to its angular momentum.
For each object, this formally corresponds to either 
a rotation of
the reference frame, or
a redefinition of
its orbit; both approaches being of course equivalent.

\subsection{Getting to this Earth-crossing subset}\label{sec:orbel_derivation}

By far 
the most frequent 
method
for
parametrising 
a generic keplerian orbit 
is
to provide a set of orbital elements:
semi-major axis $a$ (or equiv.\@ orbital period $P$), eccentricity $e$, inclination $i$, argument of perihelion $\omega$, and longitude of ascending node $\Omega$, which are constants that fully determine the orbit in a 2-body problem with point masses; the position of the orbiting body being given at some reference epoch either by its true anomaly $f$ or mean anomaly $M$.
A clear advantage of using orbital elements is that they are general and can be used for any heliocentric orbit, including main-belt and Amor asteroids for instance.

	\begin{figure}[h!]
	\begin{center}
		\includegraphics[width = .7\textwidth]{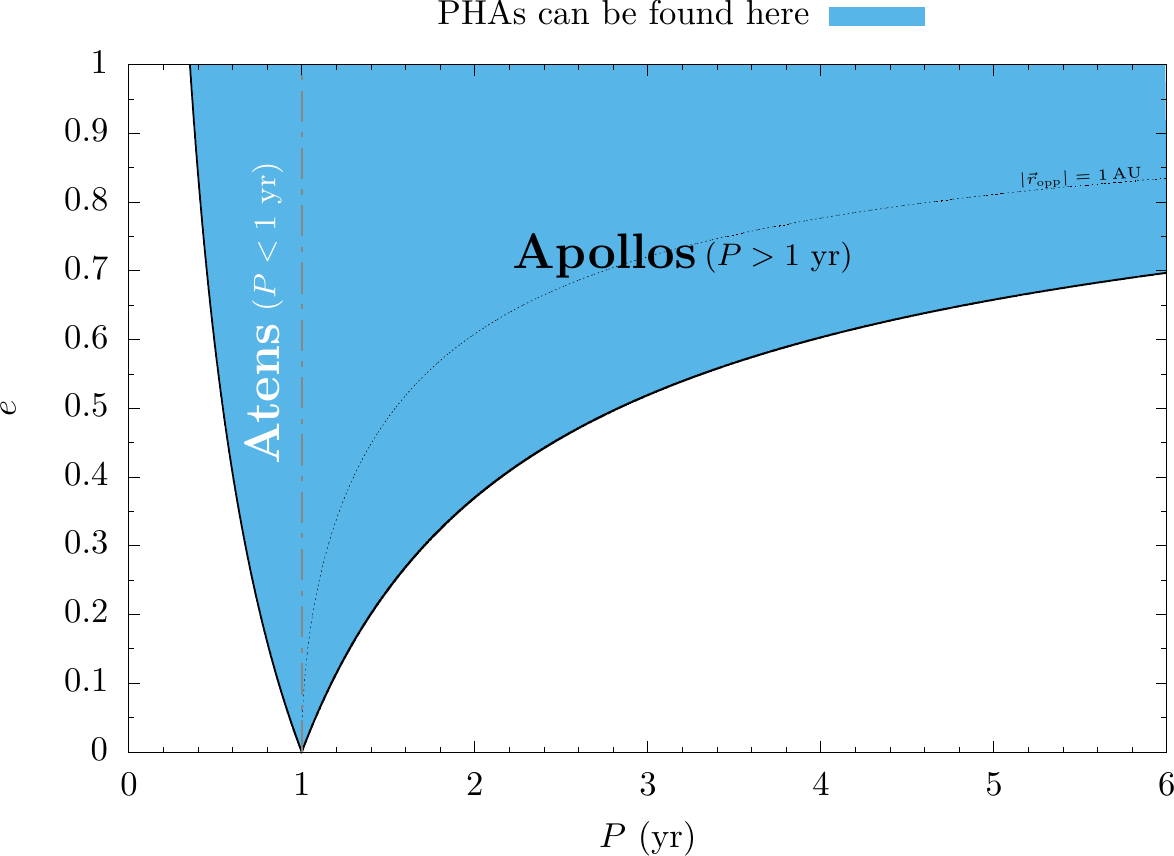}
	\end{center}
		\caption{Non-trivial region in which is satisfied the necessary but not sufficient condition on $(P,e)$ for strictly impacting elliptic orbits\hspace{1pt}---\hspace{1pt}meaning that most of these orbits do not qualify as PHAs. For those that actually are, the frequently used subdivisions ``Atens'' and ``Apollos'' are shown, as well as the locus of orbits for which $|\vec{r}_{\rm opp}| = 1$~AU (see text).}
		\label{fig:borderPHAs}
	\end{figure}

When 
one is only interested in Earth-impacting orbits however, a number of constraints must 
then 
be 
enforced.
Before that, for sheer convenience, Eq.~\eqref{eq:impactlocation} can of course be used to 
get rid of
$\Omega$, since the rotational symmetry of the circular Earth orbit gives a total freedom on its value.
Then,
for an impact to be at all possible, the asteroid orbit must 
obviously
have a perihelion smaller than 1~AU and aphelion larger than 1~AU.
This means that for each value of the period, the eccentricity must satisfy the non-trivial constraint:
\be
	e \geq \left| 1 - {\left( \frac{P}{1~{\rm yr}} \right)}^{-\frac{2}{3}} \right|;
	\label{eq:borderPHAs}
\ee
the corresponding well-known region is shown in Fig.~\ref{fig:borderPHAs}.
This is however only a necessary but not sufficient condition for a strict impact with the circular Earth orbit: if the other orbital parameters are left unconstrained, 
most of these orbits will 
indeed 
not cross the Earth due to 
their three-dimensional orientation.
A further filtering is required\hspace{1pt}---\hspace{1pt}only in 2D do they all cross;
the inclination 
being
otherwise a free parameter.

	\begin{figure}[h!]
	\begin{center}
		\includegraphics[width = .55\textwidth]{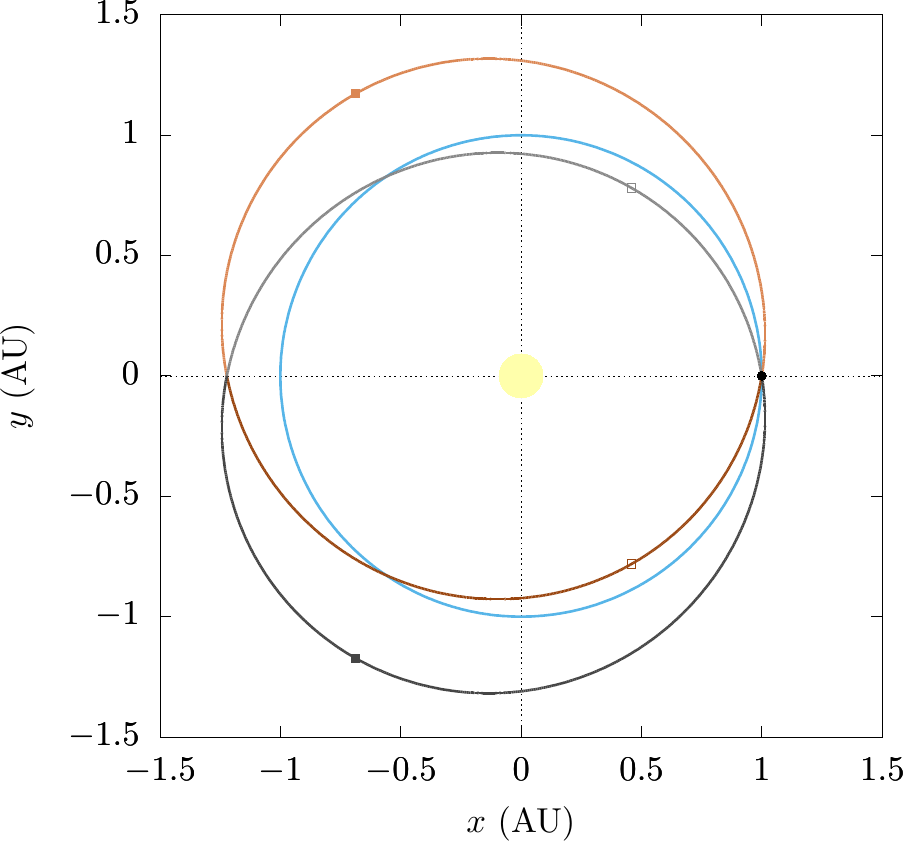}
	\end{center}
		\caption{Ecliptic-plane projection of two example impacting orbits described by exactly the same $(P,e,i)$, but oriented differently in their common orbital plane; the circular Earth orbit is also shown. The perihelion (aphelion) of each PHA orbit is shown as an open (closed) square, and the impact location $\vec{r}_I$ is indicated with a full dot. The $x$-axis is the line of nodes if $i \neq 0$; to reinforce the impression that the ecliptic is crossed in such a case, slightly different 
hues
are used on either side.}
		\label{fig:orientation_orbits}
	\end{figure}

For each such set $(P,e)$,
there exist
two 
orientations in the PHA orbital plane 
that 
lead 
to a collision with the circular Earth orbit; see Fig.~\ref{fig:orientation_orbits}.
In one case, the impact happens as the PHA is travelling from the inside of the Earth orbit to the outside (daytime impact),
and {\it vice versa} in the other (nighttime impact).
Being able to
distinguish
them
unequivocally
is 
essential,
both for astronomers and for mission analysts,
since
the properties of these two distinct orbits can be quite different\hspace{1pt}---\hspace{1pt}a salient reason being 
notably
that it is difficult to observe at low solar longitudes.\footnote{For instance, the Chelyabinsk meteor was a daytime impact, and could not be observed from ground prior to impact because it was angularly too close to the Sun~\cite{Borovicka_etal:2013,SurveysReview:2015}.}
The corresponding 
orbits
are obtained by enforcing that 
orbital parameters satisfying Eq.~\eqref{eq:borderPHAs} moreover meet the MOID condition, required
to qualify as a PHA.
A strict impact ($\textrm{MOID} = 0$) with the circular Earth orbit indeed formally means that
\be
	r^{\rm (PHA)}(f = f_I) = \frac{a(1 - {e}^2)}{1 + e \cos f_I} = 1~{\rm AU}, \label{eq:impactcondition}
\ee
where $f_I$ is the asteroid true anomaly at which the impact will eventually take place.
In the literature what is usually done is often not to write Eq.~\eqref{eq:impactcondition}, but to 
recast it as a constraint on the argument of perihelion $\omega$;
this however 
first requires choosing
explicitly
at which of the nodes
of the asteroid orbit (ascending or descending) the impact 
is to take place.\footnote{By definition, using $\omega$ demands that the ascending node of the orbit is clearly identified, since this angle is counted from this location to the perihelion. This also fixes the descending node, obviously.}
For this reason,
fictitious PHA
orbits are 
usually
arbitrarily assumed to cross the Earth orbit at their ascending node 
(thereby implying $f_I = 2\pi - \omega$):
\be
	r^{\rm (PHA)}(f = f_I) = \frac{a(1 - {e}^2)}{1 + e \cos \omega} = 1~{\rm AU}. \label{eq:impactcondition_omega}
\ee
For each pair $(a,e)$ 
that comes out of
Eq.~\eqref{eq:borderPHAs},
the equation above
obviously accepts only two solutions for $\omega$.
This is how one
frequently
distinguishes
daytime and nighttime impacts:
assuming ascending-node impacts,
the nighttime branch 
is then selected by using the value smaller than $\pi$, and the daytime branch, the one larger than $\pi$;
see again Fig.~\ref{fig:orientation_orbits}.

It is however clear that the ascending node has nothing special intrinsically, and that
the impact could equally well happen at the descending node
in 
case of 
an actual threat.
With $f_I = \pi - \omega$, nighttime impacts would
then instead
correspond to $\omega > \pi$, and daytime, to $\omega < \pi$; once again, see Fig.~\ref{fig:orientation_orbits}.

\begin{figure}[h!!]
	\begin{center}
		\includegraphics[width = .485\textwidth]{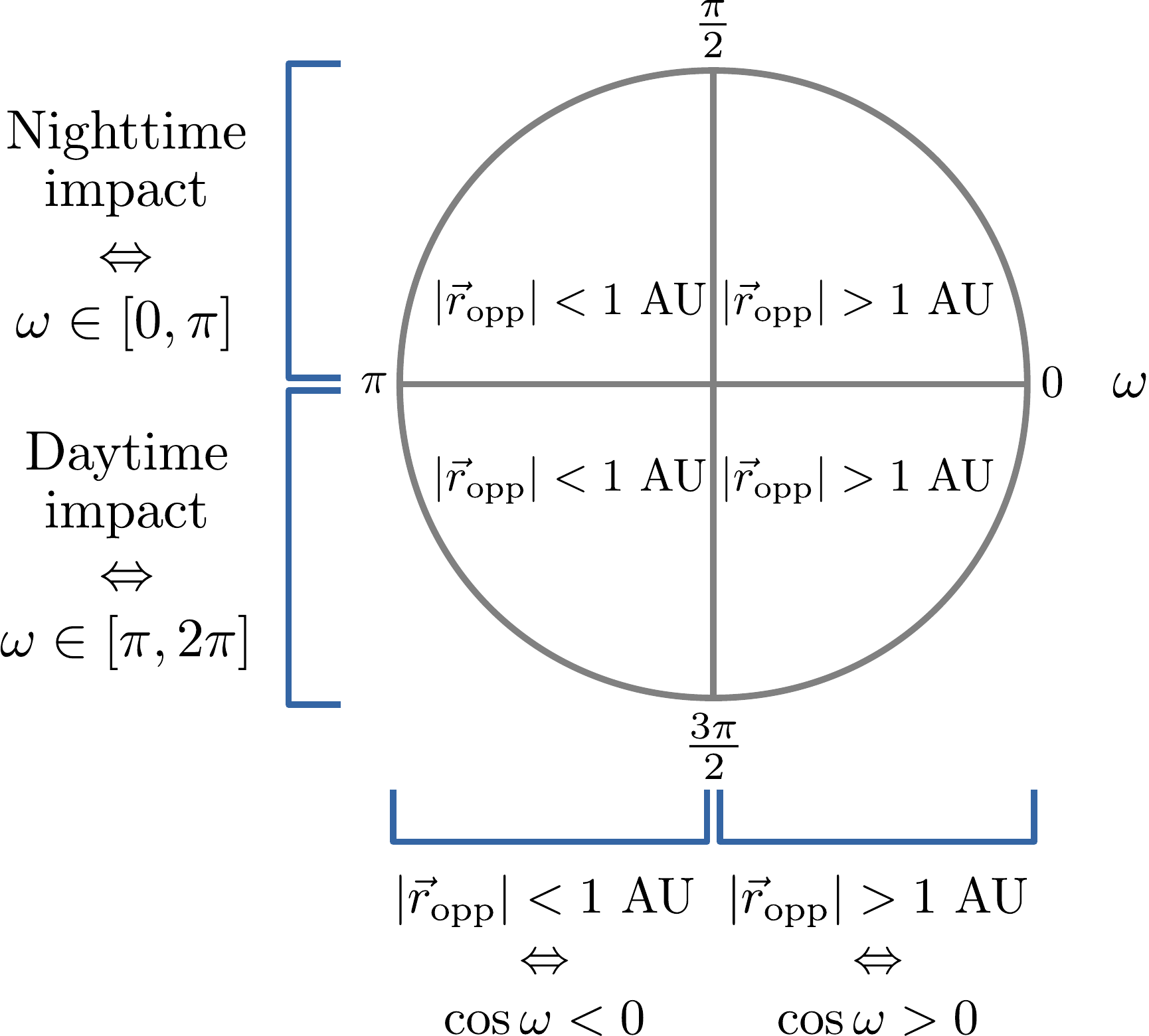}
		\hfill
		\includegraphics[width = .485\textwidth]{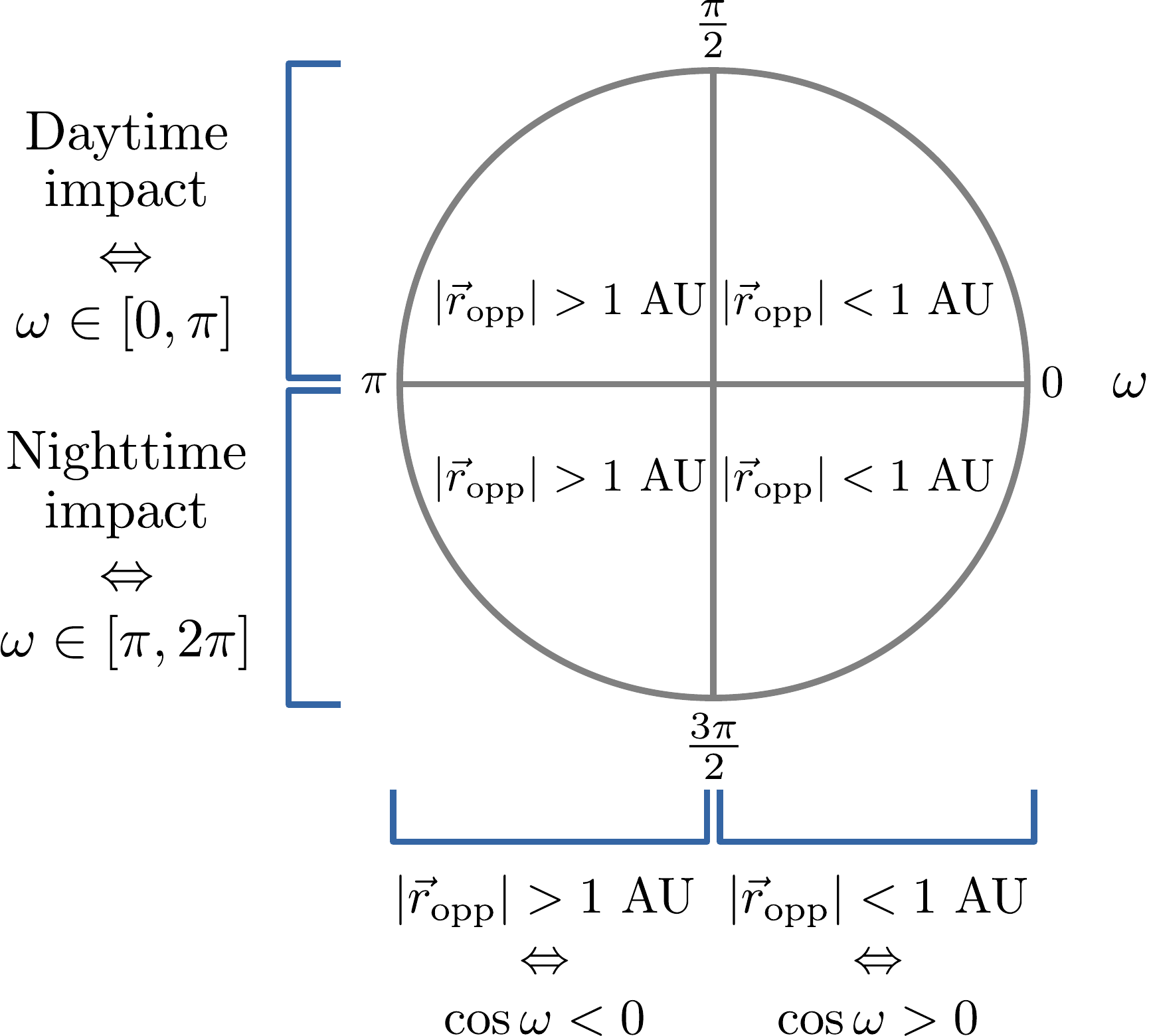}
	\end{center}
		\caption{Unit-circle representation of $\omega$, assuming either an impact at the ascending (\emph{left}), or at the descending node (\emph{right}).
		Different combinations of orbital properties can be uniquely separated in quadrants as a function of $\omega$, within each panel.
		Depending on the assumed impact node, the actual correspondence between each possible value of $\omega$ and those set of properties completely changes however.}
		\label{fig:unitcircles_omega_ASC_DES}
	\end{figure}

We summarise and compare the two possible alternatives and what they entail in Fig.~\ref{fig:unitcircles_omega_ASC_DES}.
For any given $\omega$,
we moreover indicate 
on these sketches
whether 
the other node of the 
asteroid
orbit,
here called the opposite node,
\be
	\vec{r}_{\rm opp} = -|\vec{r}_{\rm opp}| \frac{\vec{r}_I}{|\vec{r}_I|}, \label{eq:vec_ropp}
\ee
is located
inside or outside of the Earth orbit. For mission design, the opposite-node location may be especially of interest since the median inclination in the MPC \texttt{pha\_extended} database~\cite{MPC} is close to 10\textdegree{}; reaching locations
far from the nodes will therefore not always be
realistic or desirable.
The heliocentric distance $|\vec{r}_{\rm opp}|$
can be determined from this
general relation:
\be
	\frac{r^{\rm (PHA)}(f_{\rm asc})}{r^{\rm (PHA)}(f_{\rm des})} = \frac{r^{\rm (PHA)}(2\pi - \omega)}{r^{\rm (PHA)}(\pi - \omega)} = \frac{1 - e \cos\omega}{1 + e \cos{\omega}},
\ee
with $f_{\rm asc}$ and $f_{\rm des}$, the asteroid true anomaly at ascending and descending node respectively.
From Fig.~\ref{fig:unitcircles_omega_ASC_DES}, it is then clear
that the value taken by $\omega$ alone cannot 
be uniquely linked in general to a set of properties; on the contrary, it may 
actually
correspond to extremely different situations: swapping not only daytime and nighttime, but also the opposite-node location.

\subsection{Motivations for a new parametrisation}

\subsubsection{Relying on the argument of perihelion breaks a symmetry}

While using $\omega$ to distinguish daytime and nighttime impacts in Eq.~\eqref{eq:impactcondition_omega} 
necessarily
requires an arbitrary choice to be made regarding the impact node (ascending or descending), 
there is 
no physical justification
nor need
for preferring either of them.
The relative geometry of the collision indeed remains absolutely identical should the impact happen from above or from below the ecliptic plane.
It is worth realising that making 
such a
choice explicit
therefore
breaks a symmetry of the impact
problem.
The only thing that 
should matter is
whether the asteroid is moving from the inside to the outside of the Earth orbit as it crosses it, or {\it vice versa}.
Whether the node is the ascending or the descending one is
totally irrelevant in this problem.

The very use of 
the argument of perihelion
to distinguish
the orbits with daytime and nighttime impacts 
is then
ultimately
responsible for 
an unnecessary loss of clarity,
and 
introduces 
an artificial need 
for transformations
between the two arbitrary choices shown in the two panels of Fig.~\ref{fig:unitcircles_omega_ASC_DES}.

\subsubsection*{Remark: building impacting orbits from databases}

Note that Earth-impacting orbits are sometimes generated from actual orbits in PHA data\-bases or theoretical distributions of NEO orbital elements such as those of Refs.~\cite{Bottke_etal:2002, Granvik_etal:2016}, 
which are then
slightly modified
in order to achieve MOID = 0.
As is indeed expected in reality, the collisions will therefore find themselves distributed\hspace{1pt}---\hspace{1pt}essentially equally\hspace{1pt}---among the respective ascending and descending nodes.

Relying on the argument of perihelion to distinguish between daytime- and nighttime-impact orbits is then arguably even more
cumbersome to work with,
knowing that the node can 
hop from the ascending to the descending node from one PHA to the next, within the same paper:
given ranges of $\omega$ thereby constantly changing their meaning.

\subsubsection{Gaps, branches, and ill-defined behaviour}

Parametrised by means of the orbital elements, the full parameter space of impacting orbits is hardly a simple region\hspace{1pt}---\hspace{1pt}even when assuming a circular Earth orbit for simplicity; there is simply no natural relation between them.
Earth-impacting asteroid orbits are therefore essentially seen as discrete entities: one often creates a finite population, counting sometimes up to a few tens or hundreds of thousands of individual objects.
A reason for that is that taking an impacting orbit and slightly changing its orbital elements will usually not lead to another impacting orbit. They cannot be continuously deformed into one another by arbitrary jumps in parameter space.

In this standard parametrisation, one should not only deal with branches in $\omega$, but also with gaps in $(P,e)$, as we discussed during the derivation in the previous section.
There moreover exists a minimum value for the orbital period (equiv.\@ semi-major axis), below which no impact solution can be found with elliptic orbits.\footnote{Numerically, one should also decide on an upper value for $P$, which is not formally bounded.}
As for the branches, for a fixed inclination, the two possible orientations for the same $(P,e,i)$ set, distinguished using $\omega$, cannot be presented at once on the same 2D plot. 
This leads to either having to deal with 
at least
two distinct copies of the $(P,e)$ graph shown on Fig.~\ref{fig:borderPHAs}\hspace{1pt}---\hspace{1pt}each associated with a different range of values for $\omega$\hspace{1pt}---\hspace{1pt}when presenting results ({\it e.g.\@} the achieved $B$-plane deflection for each orbit with kinetic impactors~\cite{FB:2015}), or 
having to
abandon
the distinction between daytime- and nighttime-impact orbits altogether on these kinds of plots.

Finally, this parametrisation behaves particularly badly for vanishing inclinations, since $\omega$ is then obviously no longer defined. 

\bigskip

These various shortcomings are so many motivations for an alternative parametrisation. 
We now show that all of these problems can actually be avoided,
using simply as little as three dimensionless physical parameters, directly relevant to the impact problem.

\section{A new parametrisation for strict impacts: $f_I, e, i$}\label{sec:newparam}

The solution that we propose is 
to parametrise
all the conceivable elliptical 
Earth-impacting orbits
solely
by means of 
the 
inclination $i \in [0\textrm{\textdegree},180\textrm{\textdegree}]$,
the 
eccentricity $e \in [0, 1[$,
and
the 
asteroid true anomaly at impact $f_I \in [0, 2\pi[$.
The full PHA parameter space is 
indeed
very
simply given,	
without any constraints, by
\be
	\textrm{the whole set of } (f_I,e,i).
	\label{eq:fullparam_fIei}
\ee

\subsection{Earth-impacting orbits for domain of existence}

This parametrisation is very natural when the problem is considered from the asteroid viewpoint.
The asteroid true anomaly at impact $f_I$
is indeed
physically and
geometrically directly relevant both to the asteroid orbit and the impact problem: it 
simply indicates where the collision shall take place on the asteroid orbit; being nothing more than the angle from the perihelion to that point.
Since $f_I$ is an angle in the PHA orbital plane, nothing special happens if $i = 0$; the parametrisation therefore remains well-defined even in that case.

The reason why using this new parametrisation can be so simple is 
that
its domain of existence exactly matches the subspace of elliptical Earth-impacting orbits. 
For any inclination, the impact condition
\be
	r^{\rm (PHA)}(f = f_I) = \frac{a(1 - {e}^2)}{1 + e \cos f_I} = 1~{\rm AU}
	\tag{\ref*{eq:impactcondition}}
\ee
indeed necessarily admits any $(f_I,e)$ pair
as a solution because the very existence of $f_I$ 
implies that an impact shall take place, by definition.
The corresponding semi-major axis 
is then a function,
obtained by inverting the impact condition:\footnote{Rather than the orientation, it is now the semi-major axis which adapts itself to fulfil Eq.~\eqref{eq:impactcondition}.}
\be
	a(f_I,e) = \frac{1 + e \cos f_I}{1 - e^2} \ {\rm AU}.
	\label{eq:a(fI,e)}
\ee
In comparison, 
as discussed in Sec.~\ref{sec:param}, the same cannot be said about $(P,e)$:
in that case, regions devoid of solution existed because an impact cannot always be enforced for any of these pairs.
A similar conclusion holds for
a possible $(f_I, P)$ parametrisation; see App.~\ref{sec:fIP}.

Instead of
a
complicated region,
the entire set of Earth-impacting orbits
now
corresponds to
a continuous and bounded region of parameter space which is absolutely trivial and fully described by only 3 dimensionless parameters with a clear physical meaning.
Again, no filtering is required, and there is no loss of generality.

What is also important is that this region is moreover convex.
In other words,
the application of
any arbitrarily large changes to the parameters of any Earth-impacting orbit
\be
	(f_I, e, i) \rightarrow	(f_I', e', i') = (f_I + \delta f_I, \ e + \delta e, \ i + \delta i),
	\label{eq:continuousregion}
\ee
be it at once or individually, 
always leads to a valid orbit, without ever leaving the subspace of Earth-impacting orbits as long as 
each parameter remains within its respective range.
Instead of 
handling a finite
population of
distinct
objects, the new parametrisation 
thereby
opens the possibility to 
consider
the whole parameter space of PHA orbits as a continuum in $(f_I,e,i)$.

\subsection{Symmetry restored}

For distinguishing daytime- and nighttime-impact orbits,
it 
is clear from the discussion in Sec.~\ref{sec:orbel_derivation}
that
the use of $f_I$ with the impact condition given by Eq.~\eqref{eq:impactcondition} 
restores the symmetry that was broken by relying on the argument of perihelion:
daytime impacts naturally happen when $f_I < \pi$, when the collision takes place between perihelion and aphelion; whereas $f_I > \pi$ for nighttime impacts, the situation being then reversed. 

Interestingly, 
writing a general analytical relation for the position of the opposite node $\vec{r}_{\rm opp}$, introduced in Eq.~\eqref{eq:vec_ropp}, does not require choosing between ascending- and descending-node impacts either.
We
start from the following simple observation:
\be
	f_{\rm asc.} \equiv 2\pi - \omega, \quad\textrm{while}\quad f_{\rm des.} \equiv \pi - \omega = f_{\rm asc.} - \pi,\notag
\ee
which implies that
we always have
\be
	(f_I - f_{\rm opp}) \mod 2\pi = \pi = (f_{\rm opp} - f_I) \mod 2\pi
\ee
so that, using the conic equation, we get
\be
	\frac{|\vec{r}_{\rm opp}|}{|\vec{r}_I|} = \frac{1 + e \cos f_I}{1 - e \cos f_I},
	\label{eq:ropp}
\ee
independently on whether the impact point corresponds to the ascending or descending node of the PHA orbit, allowing us to remain completely general in the following.

	\begin{figure}[h!!!]
	\begin{center}
		\includegraphics[width = .485\textwidth]{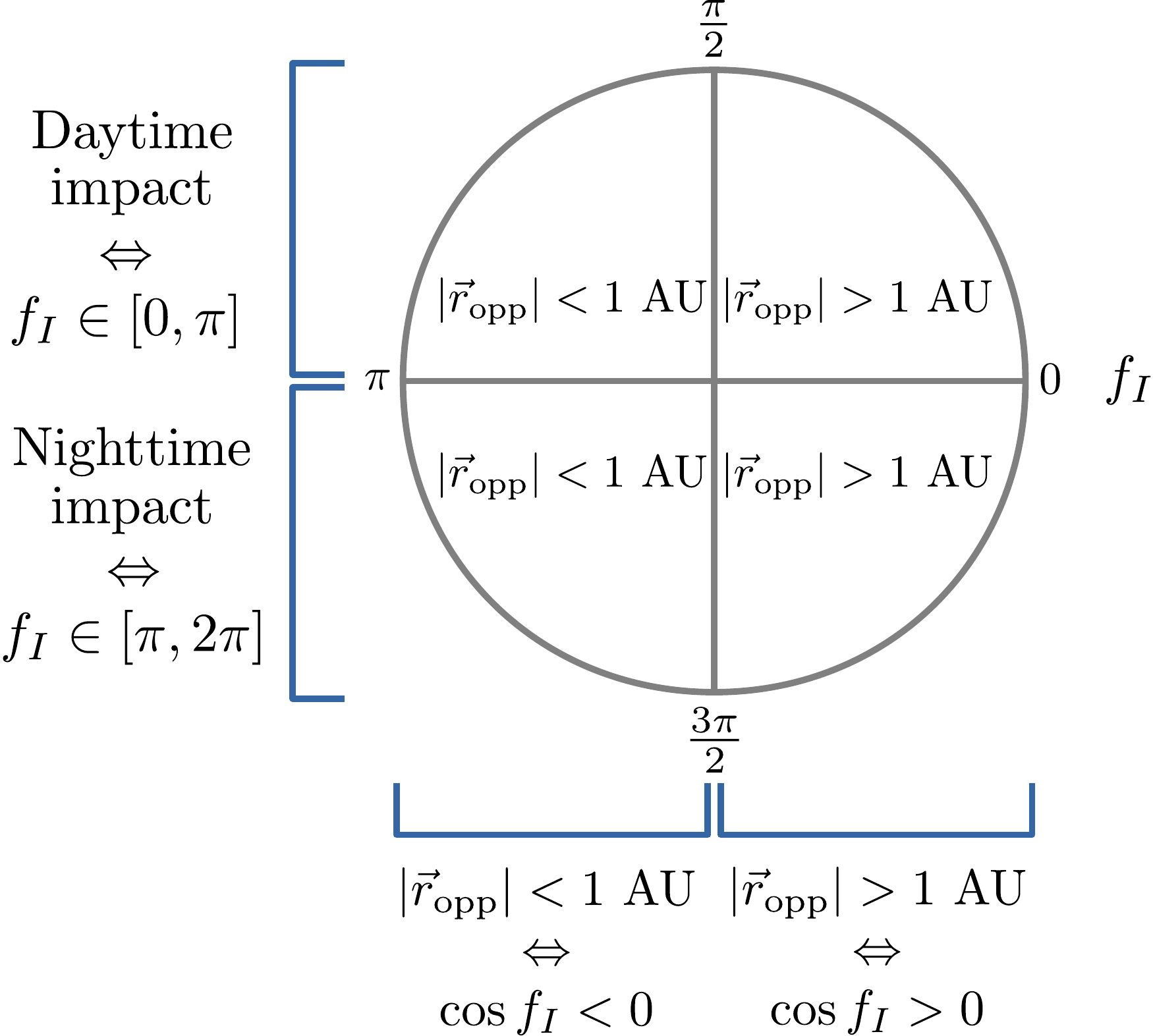}
	\end{center}
		\caption{Unit-circle representation of $f_I$; replaces both panels of Fig.~\ref{fig:unitcircles_omega_ASC_DES}.
			}
		\label{fig:scenarios_fI}
	\end{figure}

All this information is summarised in Fig.~\ref{fig:scenarios_fI}, which 
remains true in absolutely all cases.
Transformations are no longer needed.
Each quadrant in $f_I$ shall always correspond to the same combination of properties, as shown on this figure, 
no matter 
whether the impact happens at the ascending or descending node.

\bigskip

Making an explicit choice 
would 
not alter
the correspondence between 
the parametrisation
and the different orbits
and is no longer required,
but could of course still be done if needed.
For any $(f_I,e,i)$ orbit, it is for instance trivial to recover all the orbital elements:
not only the semi-major axis $a$ thanks to Eq.~\eqref{eq:a(fI,e)}, but also the argument of perihelion $\omega$ and the argument of ascending node $\Omega$.
For the two possible assumptions on the impact location $\vec{r}_I$, they read
\be
	\textrm{ascending-node impact}: \qquad  \omega = 2\pi - f_I \quad \textrm{and} \quad \Omega = 0,
\ee
whereas
\be
	\textrm{descending-node impact}: \qquad  \omega = \pi - f_I \quad \textrm{and} \quad \Omega = \pi.
\ee
A consequence of this is that the new parametrisation can moreover be considered as a simple way to generate orbital elements for Earth-impacting orbits\hspace{1pt}---\hspace{1pt}as can be useful for using existing subroutines, for instance.
It is nonetheless strongly advised to 
keep track of
$f_I$ even then. At least for distinguishing the different impacting orbits, obviously,
but for plotting as well, since this actually brings some more benefits that we now discuss.

An additional welcome byproduct of the new parametrisation is an analytical one.
Since it originates from considerations about the geometry of the impact problem, writing equations in terms of $(f_I,e,i)$ turn out to become much more elegant and intuitive\hspace{1pt}---\hspace{1pt}enough to
facilitate
analytical studies.
A number of relations written in the three-dimensional inertial ecliptic frame actually strongly remind the simple expressions which would be written in the perifocal plane.

\section{Studying the entire parameter space at once in a polar representation} \label{sec:polar}

To
better understand
how
the different
properties of PHA orbits 
behave
over the full parameter space,
we shall actually
use
a polar-plot representation
with the eccentricity as the radial variable (from 0 to 1) and the asteroid true anomaly at impact as the angular variable (from 0 to $2\pi$).
It indeed turns out that many 
quantities 
can be written as simple functions of $e\cos f_I$ and $e \sin f_I$\hspace{1pt}---\hspace{1pt}as discussed in App.~\ref{sec:maths}, and already apparent in Eq.~\eqref{eq:ropp}.
Since both daytime- and nighttime-impact orbits can now be shown together without ambiguity,
a single plot is 
sufficient to 
map all the possible orbits sharing a given inclination.

For any fixed $i$, 
a further important advantage is that there is 
moreover
an exact one-to-one correspondence between each point on the polar plot and each conceivable impacting orbit\hspace{1pt}---\hspace{1pt}the only degenerate case ($e=0$, $\forall f_I$) 
being then obviously reduced to a single point.
This tool is therefore
of interest 
not only because of its nice analytical properties 
but also
because it can be used to efficiently report any numerical results
obtained for different Earth-impacting orbits.

	\begin{figure}[h]
		\begin{center}
			\includegraphics[width = .6\textwidth]{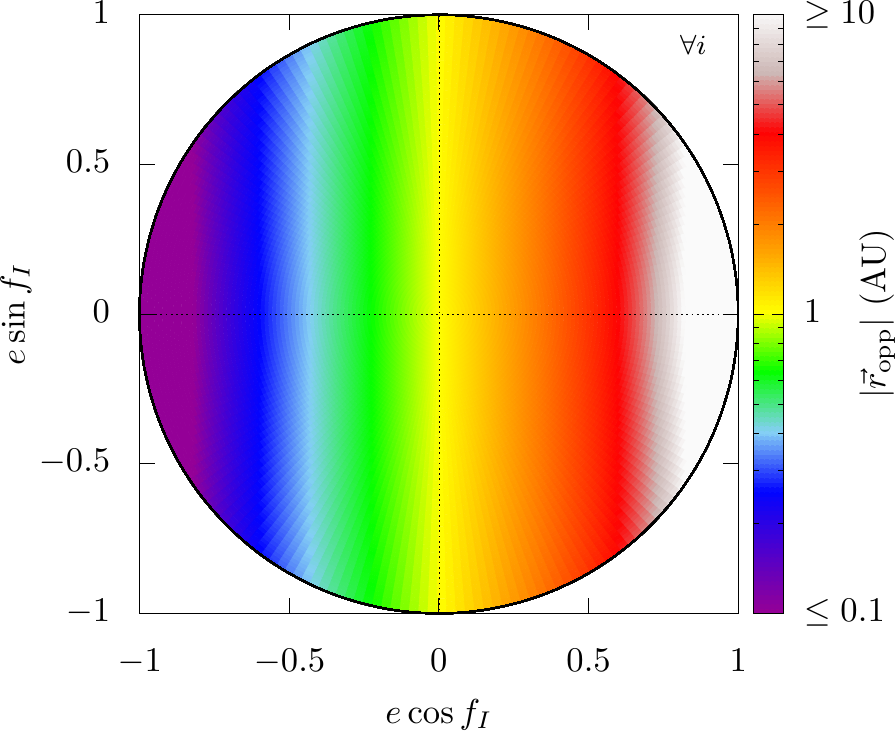}
		\end{center}
		\caption{Full PHA parameter space, in which the location of each PHA opposite node in terms of $f_I$ and $e$ is given, independently of the PHA-orbit inclination $i$, by $\vec{r}_{\rm opp} = |\vec{r}_{\rm opp}| \frac{-\vec{r}_I}{|\vec{r}_I|}$.
	}
		\label{fig:ropp_e_fI}
	\end{figure}

	A first example is to present at once for the full PHA parameter space the opposite-node location: $\vec{r}_{\rm opp}$ being independent of the inclination, a single $(f_I,e)$ plot indeed contains
	all the information for all the conceivable elliptical impacting orbits.
	We show this in Fig.~\ref{fig:ropp_e_fI}, which is a polar representation of Eq.~\eqref{eq:ropp}.
	It immediately appears that a polar-plot representation of the new parametrisation is particularly well-suited, since it efficiently highlights the dependencies.
	The different PHA orbits sharing the same $|\vec{r}_{\rm opp}|$ are indeed identified readily as vertical lines.
	Note that, on these figures, 
	the parameter space is moreover split equally between
	orbits with $|\vec{r}_{\rm opp}| < 1$~AU (negative $e\cos f_I$) and those with $|\vec{r}_{\rm opp}| > 1$~AU (positive $e\cos f_I$).\footnote{In comparison, a $(P,e)$ plot is biased towards giving more weight to orbits with $|\vec{r}_{\rm opp}| < 1$~AU, while the majority of the currently known PHA orbits~\cite{MPC} actually have $|\vec{r}_{\rm opp}| > 1$~AU; see also Sec.~\ref{sec:mpc}.}

	Additionally, please note that the role of the true anomaly at impact, which is counted clockwise,
	coincides exactly 
	with what we had in Fig.~\ref{fig:scenarios_fI}.
	The quadrants moreover exactly match.
	This therefore implies that even more information can be obtained from the polar representation, by simple comparisons:
	the different set of properties
	identified in Fig.~\ref{fig:scenarios_fI}
	indeed
	correspond to the very same quadrants in Fig.~\ref{fig:ropp_e_fI}.
	This is of course true on any such polar plot.
	Orbits with a daytime impact (positive $e\sin f_I$) or nighttime impact (negative $e\sin f_I$) are for instance identified with a simple glance.

	\bigskip

	Let us now derive
	the location of the boundary between Aten and Apollo PHAs, corresponding to PHAs having a period of one year.
	Using Eq.~\eqref{eq:impactcondition}, the condition is simply
	\be
		e^2 + e\cos f_I = 0, \qquad \textrm{so that } e = 0 \quad\textrm{ or }\quad e = -\cos f_I. \label{eq:atenapolloborder_nonpolar}
	\ee
	For the needs of our polar representation, 
	the associated locus is also easily found, simply re-write Eq.~\eqref{eq:atenapolloborder_nonpolar} as
	\be
		(e\cos f_I)^2 + (e\sin f_I)^2 + e\cos f_I = 0,
	\ee
	which in terms of $x = e \cos f_I$ and $y = e \sin f_I$ corresponds to a circle with a centre located at $(x_0 = -\frac{1}{2}, \ y_0 = 0)$ and of radius equal to $\frac{1}{2}$.

	\begin{figure}[h]
		\begin{center}
		\includegraphics[width = .6\textwidth]{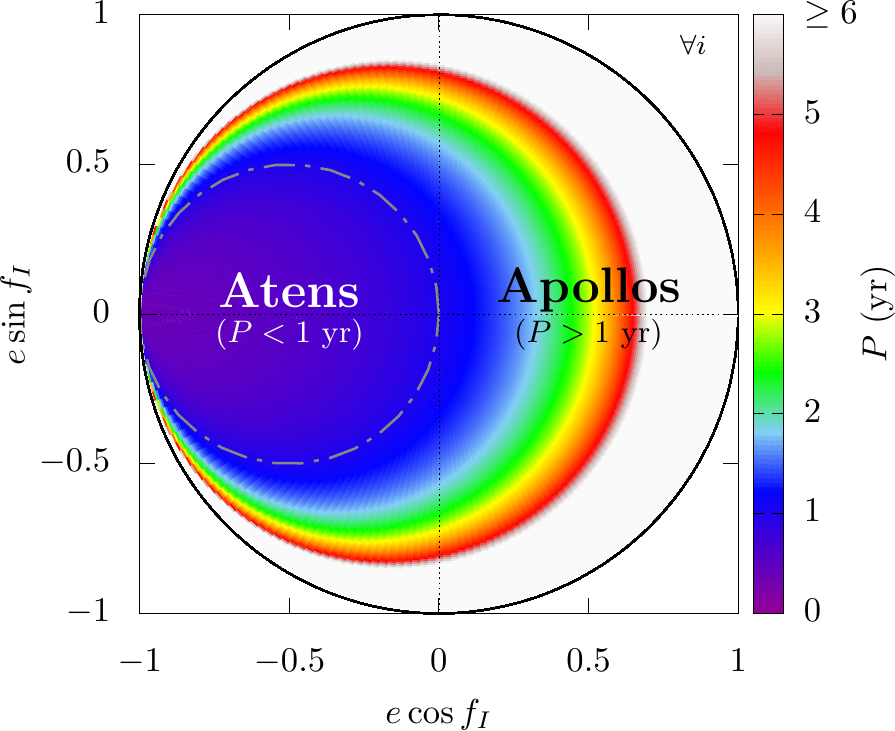}
		\end{center}
		\caption{Full PHA parameter space in a $(f_I, e)$ polar plot: representation of the orbital period. Note that the whole
		disk corresponds to PHA orbits; the colour code stops at 6 years for only for better readability.}
		\label{fig:P_fctof_fI_e_POLAR}
	\end{figure}

	In direct relation, in Fig~\ref{fig:P_fctof_fI_e_POLAR}, we also present the orbital period that corresponds to each PHA orbit in the polar-plot representation\hspace{1pt}---\hspace{1pt}which follows an interesting pattern reminiscent of the Aten--Apollo boundary.
	What is indeed striking is that each locus of orbits sharing a given orbital period coincides with a circle in the polar plot. To larger and larger values of the orbital period correspond circles of continuously increasing radius (from zero to one), and of shifting centre along the $e \cos f_I$ axis (from minus one to zero).

	As we did for the Aten--Apollo boundary, we can derive more generally the exact analytical expression for the locus
	that corresponds to any given value of the orbital period, or equivalently, to the corresponding semi-major axis.
	Equation~\eqref{eq:impactcondition} gives:
	\be
		\left( \frac{a}{1~{\rm AU}} \right) {(e \cos f_I)}^2   +    \left( \frac{a}{1~{\rm AU}} \right) {(e \sin f_I)}^2   +   (e \cos f_I)    +    \left(1 - \left( \frac{a}{1~{\rm AU}} \right)\right) = 0,
	\ee
	so that in the polar plot, each locus of PHA orbits having the same semi-major axis indeed corresponds to a circle. Written once more in terms of $x = e \cos f_I$ and $y = e \sin f_I$,
	each of these
	is determined by its centre and radius, given respectively by
	\be
		\mathscr{C} =
		\left(	x_0 = - \frac{1}{2} {\left( \frac{a}{1\hspace{2pt}{\rm AU}} \right)}^{-1}, \  y_0 = 0	\right)
		\quad
		\textrm{and}
		\quad
		\mathscr{R} =
		1 - \frac{1}{2} {\left( \frac{a}{1\hspace{2pt}{\rm AU}} \right)}^{-1}.
		\label{eq:POLAR_locus_equalperiod}
	\ee
	The equivalent result in terms of the orbital period
	of course	
	readily follows
	from
	\be
		\left( \frac{a}{1~{\rm AU}} \right)	=	{\left( \frac{P}{1~{\rm yr}} \right)}^{\frac{2}{3}}.
	\ee
	Note that we similarly discuss the aphelion and the perihelion in App.~\ref{sec:polarplot_loci_aph_peri}.

	Now, an interesting general remark regarding this formalism is that
	it is only when we consider dimensionful quantities such as $a$ (and $P$) that we explicitly introduce a physical scale,
	since the parametrisation itself relies only on dimensionless physical properties.
	Therefore, it is straightforward to generalise the results. Expressing distances in units of the relevant physical length scale 
	given by 
	$|\vec{r}_I|$,
	what precedes will hold even if $|\vec{r}_I|$ was to be given another value than 1~AU. Similarly, it is worth giving the orbital period
	in units of the period an orbit of semi-major axis $|\vec{r}_I|$ would have around its central body ({\it i.e.} one year, in this case). Here we explicitly choose 1~AU and 1~yr, but the generalisation to other values of $|\vec{r}_I|$ and $\mu_{\rm CB}$ is straightforward.

	Notice that Eq.~\eqref{eq:POLAR_locus_equalperiod} highlights the presence of a minimum orbital period for impacting elliptical orbits, or equivalently a minimum semi-major axis (for which $x_0 = -1$, $\mathscr{R}=0$):
	\be
		a^{\rm (PHA)} > 0.5 \ |\vec{r}_{I}|;
	\ee
	it indeed corresponds to the limiting case where $\vec{r}_I$ is the aphelion of an orbit with $e\rightarrow 1$.

	\subsubsection*{Practical considerations: slices in parameter space}
	
	In special cases where only PHA orbits that satisfy some specific properties would be of interest\hspace{1pt}---\hspace{1pt}{\it e.g.\@} all sharing the same $\vec{r}_{\rm opp}$ or perihelion distance, or characterised by an orbital period restricted in a certain range for example\hspace{1pt}---\hspace{1pt}parametric functions such as
	\be
		\left\{
		\begin{array}{*2{>{\displaystyle}l}p{5cm}}
			x \equiv e\cos f_I = - \frac{1}{2} {\left( \frac{P}{1~{\rm yr}} \right)}^{-\frac{2}{3}} + \left(1 - \frac{1}{2} {\left( \frac{P}{1~{\rm yr}} \right)}^{-\frac{2}{3}}\right)\cos u\\
			y \equiv e\sin f_I = \left(1 - \frac{1}{2} {\left( \frac{P}{1~{\rm yr}} \right)}^{-\frac{2}{3}}\right)\sin u
		\end{array} \right.
		\qquad \textrm{with } u \in [0,2\pi[,
		\label{eq:semimajaxis_parametricfcts_POLAR}
	\ee
	and Eqs.~(\ref{eq:aphelion_parametricfcts_POLAR}, \ref{eq:perihelion_parametricfcts_POLAR}--\ref{eq:ropp_parametricfcts_POLAR}) (see App.~\ref{sec:maths}) can be useful.
	In numerical applications in particular, such relations could be used to sample PHA orbits while controlling the steps in the orbital period over a certain range for instance, and still preserve the other advantages of the $(f_I,e,i)$ parametrisation in its polar-plot representation.

	For most uses however, and even more so when the aim is to cover the whole PHA parameter space, it is advisable to simply use $(f_I, e, i)$ directly.

\section{Some practical applications in Planetary Defence}\label{sec:furtheruses}

	\subsection{Assessing the ballistic-transfer feasibility to the opposite node}

	\begin{figure}[h]
		\begin{center}
			\includegraphics[width = .6\textwidth]{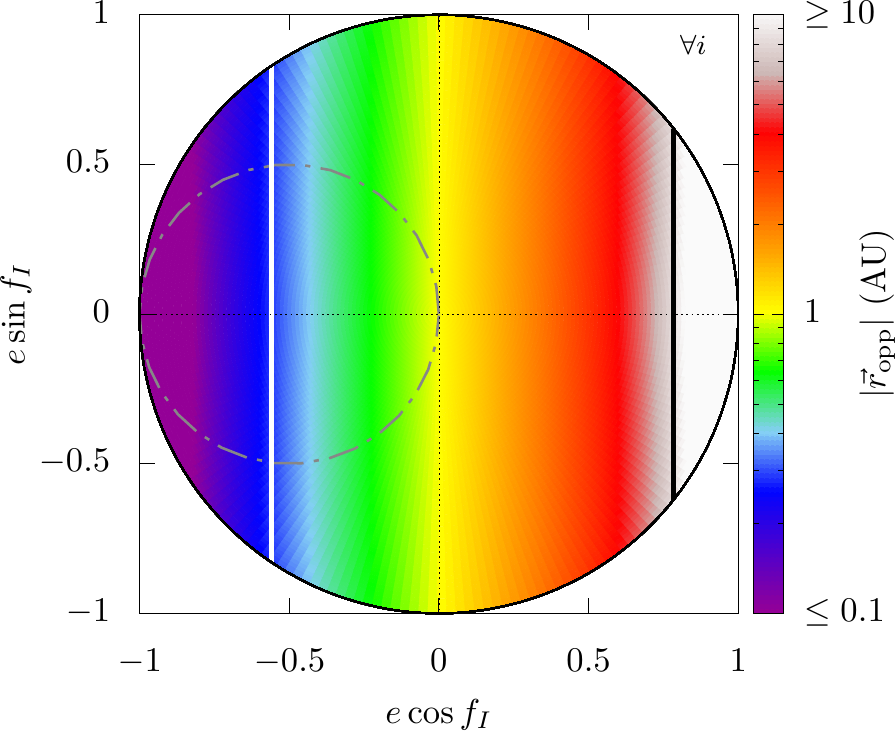}
		\end{center}
		\caption{
		Same as Fig.~\ref{fig:ropp_e_fI}, with additional overlaid information.
		The white and black lines illustrate the 
		feasibility conditions
		for a transfer to $\vec{r}_{\rm opp}$ in the case ${\max}(|\vec{v}_{\infty {\rm SC},L}|) = 10$~km/s,
		for which one finds respectively $q^{\rm (SC)}_{\rm min} = 0.283$~AU and $Q^{\rm (SC)}_{\rm max} = 8.268$~AU.
		The dashed circle indicates the Aten--Apollo boundary.
	}
		\label{fig:3dpolar_ropp_e_fI}
	\end{figure}

	Together with the $\vec{r}_{\rm opp}$ information,
	what
	would 
	be
	interesting to 
	identify
	for instance,
	in a planetary-defence context,
	is
	for which
	PHAs
	a 
	purely ballistic transfer to the opposite node would be at all possible, 
	with a given launcher.
	To determine this, we can simply decide which is the maximum hyperbolic excess velocity at launch ${\max}(|\vec{v}_{\infty {\rm SC},L}|)$ that we would be willing to consider.

	The result is overlaid on Fig.~\ref{fig:3dpolar_ropp_e_fI}  (see App.~\ref{sec:ropp_and_feasibility} for a simple derivation).
	PHAs whose $\vec{r}_{\rm opp}$ correspond to the smallest reachable perihelion that could be reached ballistically in the ecliptic plane by the spacecraft for a given  ${\max}(|\vec{v}_{\infty {\rm SC},L}|)$ are conveniently found to lie on a vertical line (shown in white) when using the polar-plot representation; the same is true for those corresponding to the largest aphelion (shown in black).
	The opposite node of PHAs found respectively to the left and to the right of these white and black lines in parameter space cannot be reached with a ballistic transfer orbit in the ecliptic plane.
	Note that, as the locations of these difficult regions are determined analytically, this assessment can be done for any launcher.

\begin{figure}[h!]
	\begin{center}
		\includegraphics[height = 7.625cm, trim = 0 0 45 0, clip]{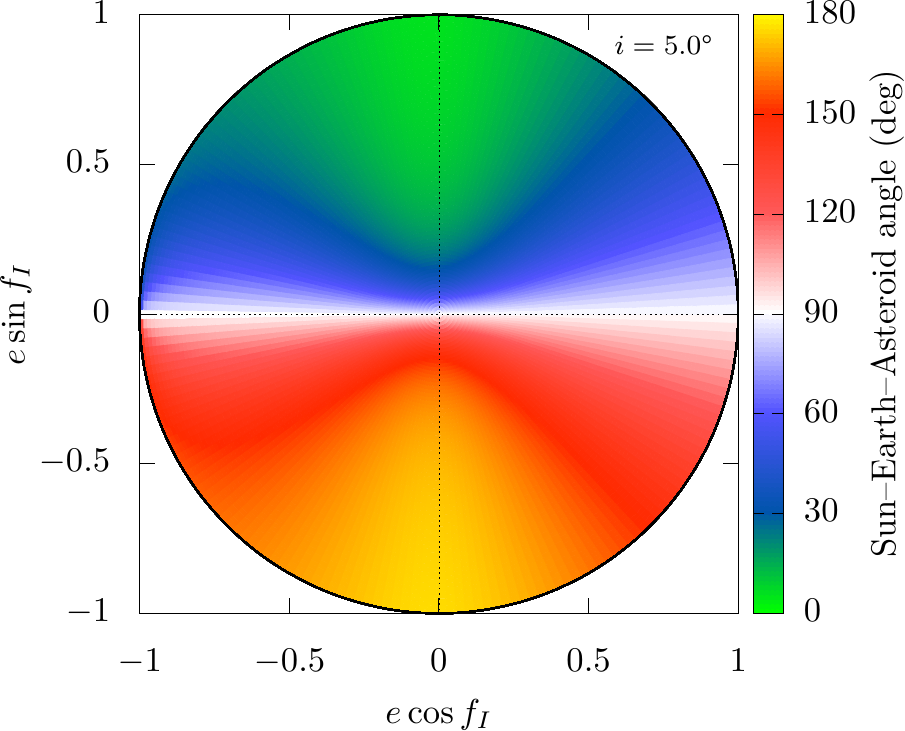}
		\includegraphics[height = 7.625cm, trim = 35 0 0 0, clip]{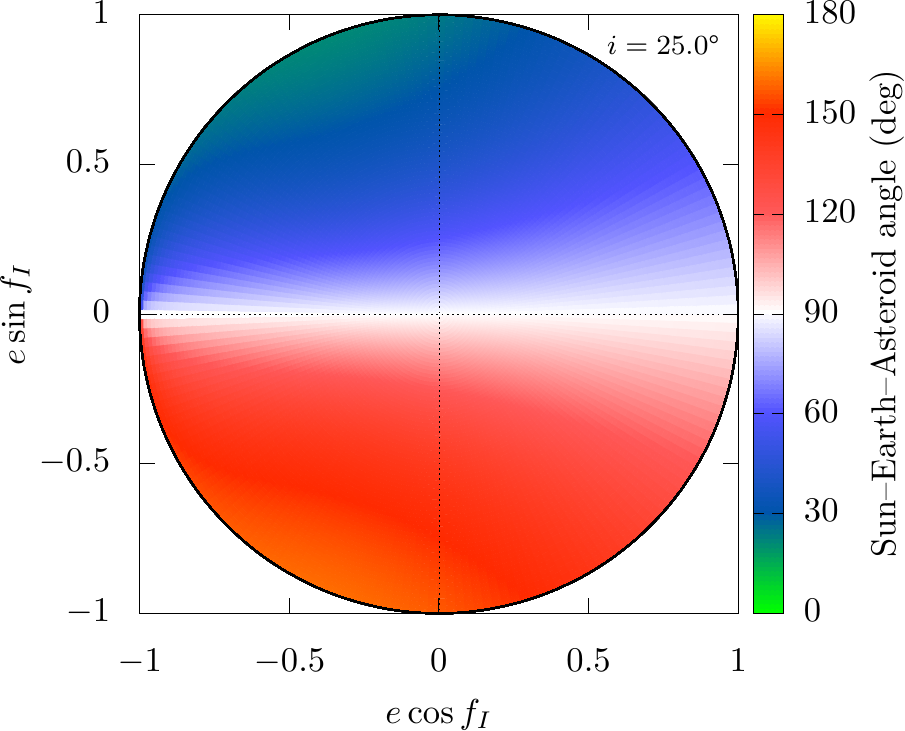}
	\end{center}
		\caption{Sun--Earth--Asteroid angle before impact, for two values of the PHA-orbit inclination.}
		\label{fig:IPA_i5deg_phiSunPolar}
	\end{figure}

	Taking into account the necessarily limited launcher performance, it is intuitively clear the optimal deflection location with a kinetic impactor cannot be expected to be the asteroid perihelion in general, simply because 
	the required trade off to reach the perihelion may be too unfavourable; see also {\it e.g.\@} Refs.~\cite{Carusi_etal:2008, VasileColombo:2008, FB:2015, APJS:defl}.
	As touched upon earlier, since it is difficult to venture far out of the ecliptic plane, the vicinities of the two nodes $\vec{r}_I$ and $\vec{r}_{\rm opp}$ actually stand out as interesting locations, especially when the asteroid-orbit inclination becomes sizeable.
	Knowing when $\vec{r}_{\rm opp}$ cannot easily be reached
	can
	therefore 
	already help identify regions in the PHA orbital parameter space for which further difficulties might be expected to arise with a kinetic impactor:
	when
	the illumination conditions 
	are
	too unfavourable in the vicinity of $\vec{r}_I$ in particular\hspace{1pt}---\hspace{1pt}{\it i.e.\@} for daytime impacts. 
	To quickly assess this, we can for instance calculate the Sun--Earth--Asteroid angle before impact, as the asteroid enters the sphere of influence of the Earth,
	\be
		{\rm \widehat{SEA}}_I = \acos\left(\frac{v_{A,I,x}}{\left|\vec{v}_{\infty A,I}\right|} \right), \qquad  \textrm{with } \left|\vec{v}_{\infty A,I}\right| = \sqrt{ {v_{A,I,x}}^2 + {(v_{A,I,y} - v_{E})}^2 + {v_{A,I,z}}^2 },
		\label{eq:phiSun_ipa}
	\ee
	where $v_E$ is the norm of the velocity on the circular Earth orbit, $\vec{v}_{\infty A,I}$ is the hyperbolic excess velocity of the asteroid with respect to the Earth, while simple analytical relations for the components of the asteroid heliocentric velocity at impact $\vec{v}_{A,I}$ are given in App.~\ref{sec:maths}.
	This angle is shown for two values of the inclination in Fig.~\ref{fig:IPA_i5deg_phiSunPolar}. Note that the daytime-impact ($f_I < \pi$) and nighttime-impact ($f_I > \pi$) distinction appears clearly on these plots; notice also the symmetry between these cases within each panel.
	Comparing now with Fig.~\ref{fig:3dpolar_ropp_e_fI}, we can quickly identify which are the PHA orbits for which a deflection in the vicinity of either node will be difficult (or impossible) with a kinetic impactor, even when assuming ${\max}(|\vec{v}_{\infty {\rm SC},L}|) = 10$~km/s: around $\vec{r}_I$, because of the illumination conditions (unfavourable phase angle and solar aspect angle); around $\vec{r}_{\rm opp}$, because of the limited launcher performance.
	It is therefore clear that, should an Earth-impacting asteroid ever be found on such an orbit with a sizeable inclination, finding a solution could be considerably challenging. An example of a non-impacting but very similar PHA is \mbox{2014 JO25}~\cite{2014JO25} (see Table~\ref{table:2014JO25} in the next section); having a 
an irregular shape
and a size of about 850~m,
it actually made an Earth fly-by, approaching our planet within less than five times the lunar distance, in April 2017\hspace{1pt}---\hspace{1pt}that is, less than three years after its discovery.\footnote{Note that this particular object will not come close to the Earth again before more than 400 years.}

	This kind of information
	is actually relevant even when more involved trajectory designs are being considered, since the absence of solution in the simplest case can affect the type of allowed solutions in those cases as well: leading to longer transfer times for instance, or a poorer performance.

\subsection{IAU Minor Planet Center: PHA database} \label{sec:mpc}

	It is instructive to discuss how the new parametrisation could be related to actual, not necessarily impacting, PHA orbits.
	This is obviously an approximation.
	Its usefulness shall of course depend on how well it can be considered to hold for each object.
	It is clear that the smaller the MOID, the more accurate the description of the actual orbit will be.

	For both impacts and fairly close shaves, this could conveniently convey a lot of information.
	At the very least, it would immediately tell whether the asteroid approaches from daytime or nighttime. As we just saw, it would moreover be easy to assess the illumination conditions at impact and estimate the opposite node location and determine whether it is reachable or not with a given launcher, for instance.
	Eventually,
	if the result of mitigation studies are shown using this parametrisation, one could moreover quickly identify which kind of mitigation mission would be best suited for any PHA with a small MOID; a step in this direction is done in Ref.~\cite{APJS:defl}.

	\begin{figure}[h!!]
		\begin{center}
			\includegraphics[width = .6\textwidth]{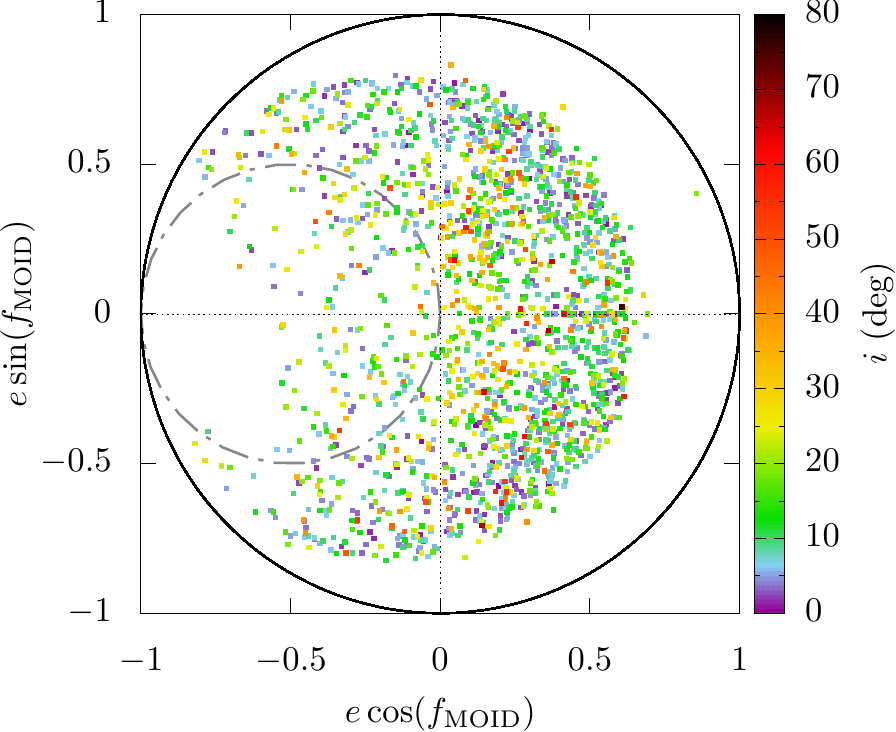}
		\end{center}
		\caption{MPC PHA database represented in a polar plot; $f_{\rm MOID}$ being used as a proxy for $f_I$.
			All objects can be shown at once, and the distinction between daytime- and nighttime-impact orbits is made clear.}
		\label{fig:all_PHAsfMOIDnominal_e_SpicelikeEarth}
	\end{figure}

	When the asteroid is not actually impacting, the concept of $f_I$ can for instance be approximated by $f_{\rm MOID}$: the asteroid true anomaly at which the MOID with the Earth orbit is realised.
	Figure~\ref{fig:all_PHAsfMOIDnominal_e_SpicelikeEarth} was made using the \texttt{pha\_extended} database provided 
	by the International Astronomical Union's Minor Planet Center~\cite{MPC}.
	Note that the object with the largest inclination is systematically shown on top when an overlap happens, to remind the reader that the inclination of known PHA orbits can be sizeable (again, the median inclination in the sample is close to 10\textdegree{}).

	\begin{table}[h!!]
		\centering
		\begin{tabular}{l r r} 
			\multicolumn{2}{c}{2014 JO25} \\[1ex]
			\hline\hline
			\multicolumn{2}{c}{Minor Planet Center values}\\
			\hline
			$a$		&2.0682656 AU\\
			$e$		&0.8854329\\
			$i$		&25.26993\textdegree{}\\
			$\omega$	&49.57126\textdegree{}\\
			$\Omega$	&30.65278\textdegree{}\\
			\hline
			$H$		&17.8 mag\\

			\hline
			\hline
			\\[-0.25ex]
			\hline
			\hline

			\multicolumn{2}{c}{True anomaly at MOID/``Impact''}\\
			\hline

			$f_{\rm MOID}$	&2.251 (129.0\textdegree{})\\

			Using Eq.~\eqref{eq:fI_of_a}	&2.246 (128.7\textdegree{})\\
			\hline
			\hline
		\end{tabular}
		\caption{\emph{Top panel}: central values at epoch 2458000.5 and absolute magnitude reported by the MPC~\cite{MPC}.
		\emph{Bottom panel:} asteroid true anomaly at MOID in a simple model, and an estimate assuming an impact.}
		\label{table:2014JO25}
	\end{table}

	For each object in the database, $f_{\rm MOID}$ was determined by means of numerical optimisations.
	As an example, in Table~\ref{table:2014JO25}, we show the orbital elements and the corresponding $f_{\rm MOID}$ obtained for the asteroid 2014 JO25 within a simple model.\footnote{More precisely, here we used the Spice kernel DE405 to get the Earth orbital elements at
	epoch mjd2000.0 (modified to have an orbital period of 1~yr), and assumed simple keplerian motions for the Earth and the PHAs.}
	The corresponding MOID that we obtained for that specific PHA is 0.01142~AU, whereas NEODyS for instance quotes 0.01184~AU~\cite{NEODyS}.
	Because $f_{\rm MOID}$ is smaller than $\pi$ (180\textdegree) for that object, we immediately know that its close approach is a daytime one; it is clear without having to plot its orbit.

\bigskip

For completeness, if the MOID can be assumed to be vanishing in order to make a quick and rougher assessment, an alternative to calculating $f_{\rm MOID}$ could be to simply
invert Eq.~\eqref{eq:impactcondition}:
\be
	f_I = \mathrm{acos}\left( \frac{1}{e} \left(\frac{a}{{\textrm {\scriptsize 1\hspace{2pt}AU}}} (1 - {e}^2) - 1\right) \right)
	\quad {\rm or}	\quad
	f_I = -\mathrm{acos}\left( \frac{1}{e} \left(\frac{a}{{\textrm {\scriptsize 1\hspace{2pt}AU}}} (1 - {e}^2) - 1\right) \right) + 2\pi,
	\label{eq:fI_of_a}
\ee
	which is obviously exact for a strictly Earth-impacting orbit. There necessarily are two possible solutions for $f_I$ since, as discussed before, each $(P,e)$ pair can correspond to two distinct branches. When provided with the orbital elements of a given PHA and no information regarding the impact location,
	one would then need the $\omega$ information to determine which of the two solutions for $f_I$ should be used.
	Beware that the resulting approximate orbit obtained with this a simpler method tend to be less satisfactory than when using $f_{\rm MOID}$. The orientation of the orbit, and therefore the conditions around close approach in particular, might not always be well preserved compared to the actual PHA trajectory.

\cleardoublepage

\section{Conclusions and perspectives}

The aim of this paper was to present
a new parametrisation suitable for studies 
performed over
the entire phase-space of Earth-impacting orbits,
such as preliminary mission-design assessments in the context of planetary defence.
To make clear why this might be needed, we first provided a reminder of the well-known complications that have to be faced when relying on 
keplerian orbital elements for studying strict impacts. It was then emphasised that 
these 
shortcomings 
are not intrinsic to the problem at hand, but merely artefacts
of 
this usual parametrisation,
and that they can therefore be avoided.
A simple alternative parametrisation, directly motivated by Earth-impacting asteroid orbits and showing none of these artificial issues, was then put forward and shown to bear many benefits.

These benefits include the realisation that the domain of existence of this new parametrisation exactly coincides with the whole subset of Earth-impacting orbits, and that the corresponding region in parameter space actually becomes trivial and convex.
Rather than considering a finite sample of distinct objects, this entire set of conceivable impacting orbits now corresponds to a continuum.
Because the new parametrisation restores a symmetry which is usually broken, 
the relative impact geometry 
is
also
made much clearer.
All this is achieved without any loss in generality, and without introducing any further assumption beyond the usual physically motivated approximations already found frequently in the planetary-defence literature. 
This new approach
actually relies on only three physical parameters, which are dimensionless, bounded, and directly relevant to the impact problem.
For any orbit, it moreover remains trivial to recover the corresponding keplerian elements if needed. 

An added advantage discussed here is
that
mathematical relations themselves become more elegant and simpler, thereby facilitating analytical studies.
As hinted by the dependencies of a number of orbital properties,
it is advantageous to present the new parametrisation in the form of polar plots. 
These allow for a visualisation of the whole parameter space at once, since there actually is a one-to-one correspondence between each point and each conceivable impacting orbit.
They therefore represent a powerful mapping tool for any kind of analytical property or numerical result that would be obtained for all the possible PHA orbits.
In particular,
the dependencies are actually made much clearer.
The polar plot was shown to be divided in quadrants, separating the entire parameter space of impacting orbits without any ambiguity based on orbital properties: not only between daytime and nighttime impacts, but also depending on whether the opposite node lies inside or outside of the Earth orbit.

\bigskip

Finally, we stress that this parametrisation is not only well-suited for 
studying various intrinsic PHA-orbit properties in the parameter space
and
assessing the feasibility of simple transfers.
More interestingly, it may indeed also be used to efficiently summarise the performance of any mitigation technique over the entire PHA parameter space.
For instance, in a separate paper~\cite{APJS:defl}, we study the results of optimised ballistic kinetic-impactor mitigation missions and show that using the new parametrisation enables a better understanding of the different mission types that can be found in various parts of the PHA parameter space. The existence of such broad regions had already been shown in Ref.~\cite{FB:2015} but,
using a parametrisation in terms of the orbital elements,
the interpretation of
these results on physical grounds
and
their identification in parameter-space
actually
remained
quite difficult for the most part.

Because a number of dependencies are 
clearly highlighted
with the new parametrisation, one can actually better understand the presence of the different mission types in different parts of the parameter space.
To some extent, even the optimisation of deflection missions to any Earth-impacting asteroid can be addressed analytically in simplified settings.

\section*{Acknowledgements}

The work reported here was done in the context of a research fellowship in the Mission Analysis section at the European Space Operations Centre (ESOC).
We thank both Michael Khan and Johannes Schoenmaekers 
for a number of interesting discussions and comments.
This research has made use of data and/or services provided by the International Astronomical Union's Minor Planet Center. 

\appendix

\appendix

\section{On a possible $(f_I, P)$ PHA-orbit parametrisation}\label{sec:fIP}

\begin{figure}[h!!]
		\begin{center}
			\includegraphics[width = .7\textwidth]{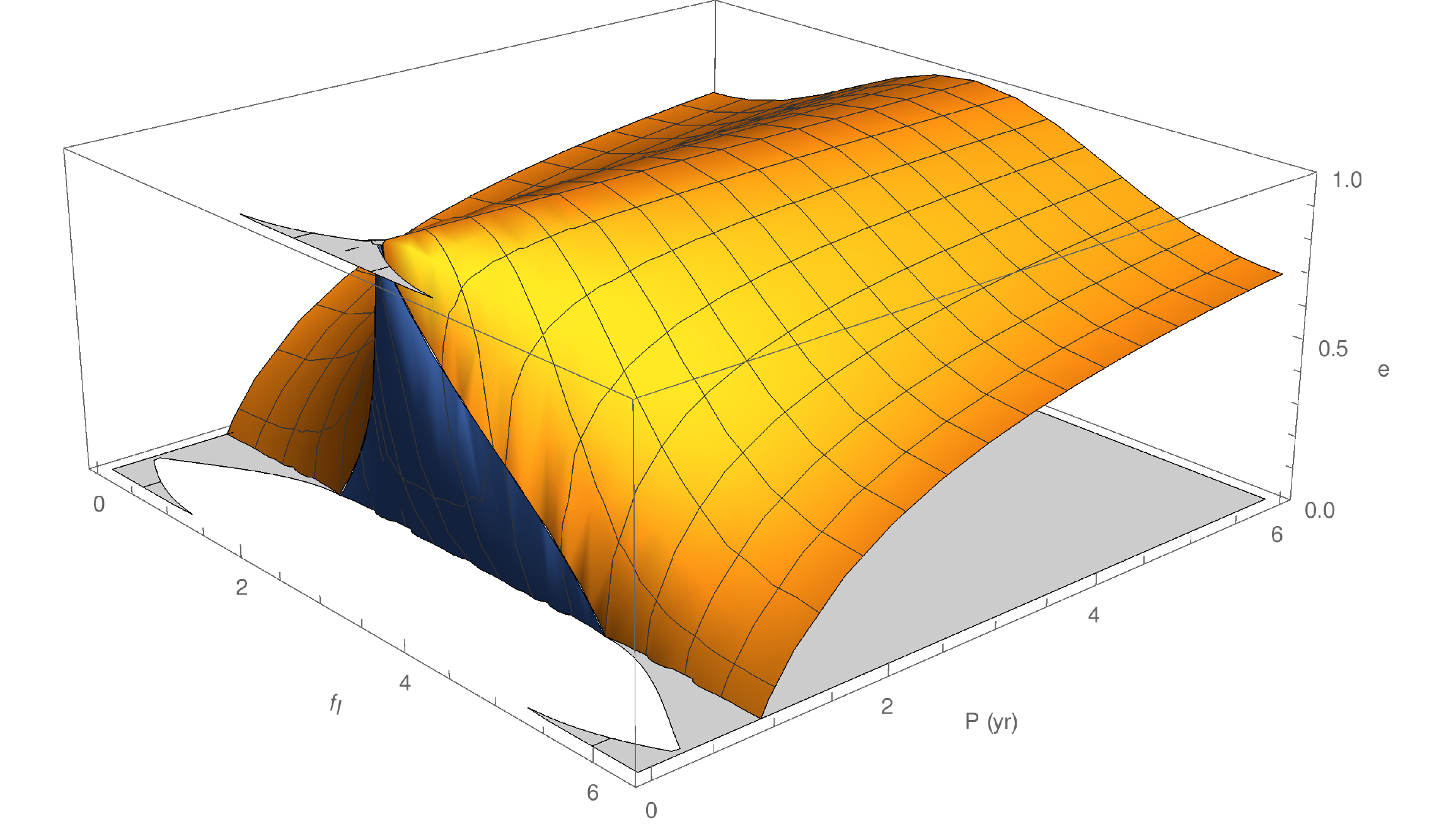}
		\end{center}
		\caption{$(f_I,P)$ PHA-orbit parametrisation, for fixed $i$. In addition to the presence of gaps, there is an ambiguity as $f_I$ and $P$ cannot uniquely determine the PHA-orbit eccentricity.}
		\label{fig:PHAs_in_fI_P_plot}
	\end{figure}

\begin{figure}[h!!]
		\begin{center}
			\includegraphics[width = .55\textwidth]{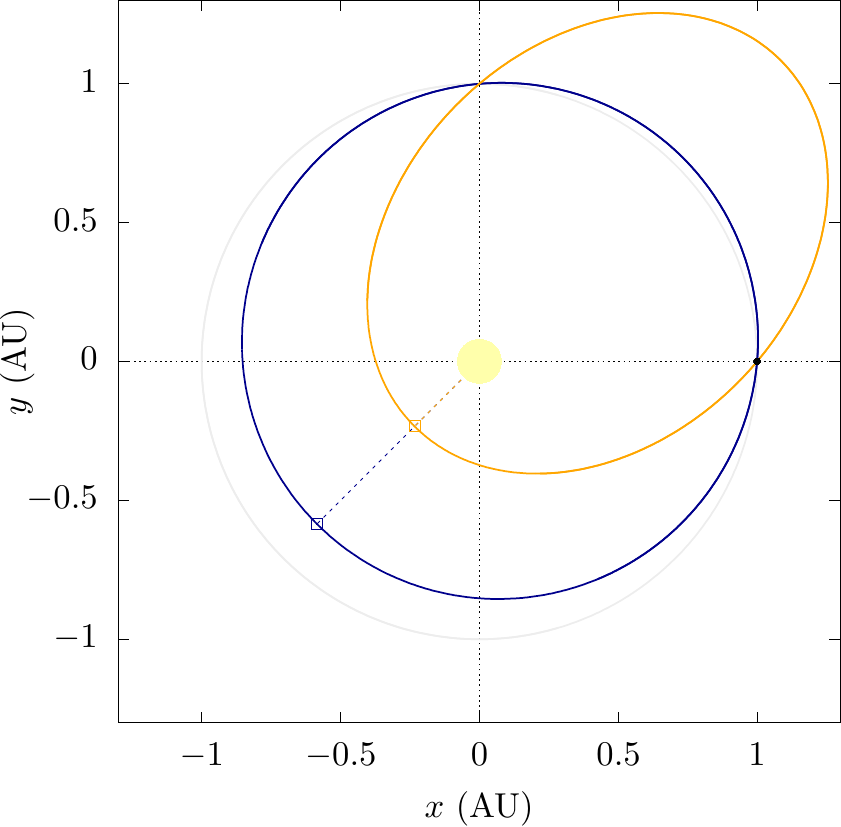}
		\end{center}
		\caption{Illustration of two PHA orbits with clearly different $e$, but actually described by exactly the same ($f_I = \frac{3}{4}\pi$, $P = 0.9$~yr) couple\hspace{1pt}---\hspace{1pt}for simplicity $i=0$\textdegree{} in this example.}
		\label{fig:comparing_same_fI_and_P_PHAs_with_diff_ecc}
	\end{figure}

	We can identify a number of downsides associated with the use of a $(f_I, P)$, illustrated in Fig.~\ref{fig:PHAs_in_fI_P_plot}, rather than of the $(f_I,e)$ parametrisation introduced in this paper:
	\begin{itemize}
		\item $(f_I,P)$ does not uniquely corresponds to a given PHA eccentricity; therefore a given $(f_I,P)$ couple can correspond very different PHA orbits, even for fixed inclination (at $P < 1$~yr); an example is given in Fig.~\ref{fig:comparing_same_fI_and_P_PHAs_with_diff_ecc};
		\item We have to choose a maximum value for $P$, while with $(f_I,e)$ we have the full PHA parameter space with the intervals $[0, 2\pi]$ and $[0,1]$, respectively;
		\item There are couples of points in $(f_I,P)$ which do not correspond to possible PHA orbits, {\it i.e.\@} there are gaps.
	\end{itemize}
	Moreover, we can anticipate that quantities that do not depend on the eccentricity, but on the semi-major axis will be more complicated in $(f_I,P)$ than either in the old $(P,e)$ or in the new $(f_I,e)$ parametrisations.

\section{Further mathematical properties of the Earth-impacting orbits} \label{sec:maths}

\subsection{Geometry at the nodes}

	Let us set ourselves explicitly in the reference frame presented at the beginning of Sec.~\ref{sec:param}, 
	with the $xy$-plane identified with the ecliptic, $\vec{r}_I$ along the $x$ axis, and the $z$-axis in the direction of the Earth angular momentum, so that the Earth velocity at $\vec{r}_I$ is $v_E \vec{e}_{y}$, with 
	\be
		v_E = \sqrt{\frac{\mu_{\rm Sun}}{\rm 1\hspace{2pt}AU}}.
	\ee
	In that external frame, the position vector of any PHA on its orbit actually simplifies into
	\be
	\begin{split}
		\vec{r}(f) 
					&=  r \cos(f - f_I) \vec{e}_x + r \sin(f - f_I)\big[\cos i\ \vec{e}_y	\pm \sin i\ \vec{e}_z\big],
	\end{split}
	\ee
	where $r=r(a(f_I,e),e,f)$ is given by the conic equation, while the positive (resp. negative) sign corresponds to an impact at the ascending (resp. descending) node of the PHA orbit.
	Anywhere on this orbit, the polar unit vectors $\vec{e}_r(f) = \frac{\partial}{\partial r}\vec{r}$ and $\vec{e}_f(f) = \frac{1}{r}\frac{\partial}{\partial f}\vec{r}$ simply read
	\be
	\begin{split}
		\vec{e}_r(f)  
					  &=  \cos(f - f_I) \vec{e}_x + \sin(f - f_I)\big[\cos i\ \vec{e}_y	\pm \sin i\ \vec{e}_z\big],
		\label{eq:e_r(f)}
	\end{split}
	\ee
	so that $\vec{e}_r(f_I) = \vec{e}_x$ and $\vec{e}_r(f_{\rm opp}) = -\vec{e}_x$, and 
	\be
	\begin{split}
		\vec{e}_f(f)
					&=  \cos(f - f_I)\big[\cos i\ \vec{e}_y	\pm \sin i\ \vec{e}_z\big] -\sin(f - f_I) \vec{e}_x,
		\label{eq:e_f(f)}
	\end{split}
	\ee
	giving for instance $\vec{e}_f(f_I) = \cos i\ \vec{e}_y	\pm \sin i\ \vec{e}_z$ and $\vec{e}_f(f_{\rm opp}) = -\cos i\ \vec{e}_y	\mp \sin i\ \vec{e}_z$.

	Exploiting the relations \eqref{eq:e_r(f)} and \eqref{eq:e_f(f)}, the expressions for the velocity of an Earth-impacting asteroid on any point of its orbit in our reference frame ($\vec{e}_x,\vec{e}_y,\vec{e}_z$) can then be easily written, from their well-known relations in the orbital frame ($\vec{e}_r,\vec{e}_f$):
	\be
	\begin{split}
		v_{A,r}(f) &= \sqrt{\frac{\mu_{\rm Sun}}{p}} e \sin f \\
		v_{A,f}(f) &= \sqrt{\frac{\mu_{\rm Sun}}{p}} (1 + e \cos f)\\
		\left|\vec{v}_{A}(f)\right| &= \sqrt{\frac{\mu_{\rm Sun}}{p}} \sqrt{1 + 2 e \cos f + e^2},
	\end{split}
	\ee
	where one shall replace $p = \left(1+e\cos(f_I)\right)$~AU, as is given by the impact condition~\eqref{eq:impactcondition}.

\subsubsection{Heliocentric PHA velocity at the nodes}\label{sec:velocity_at_nodes}

Our choice of reference frame leads to very simple expressions for PHA heliocentric velocity at both nodes. Here again, a parametrisation in terms of $f_I$ and $e$, for fixed $i$ will be useful. 

\subsubsection*{At the impact location}

	\begin{figure}[h!!]
	\begin{center}
		\includegraphics[width = .6\textwidth]{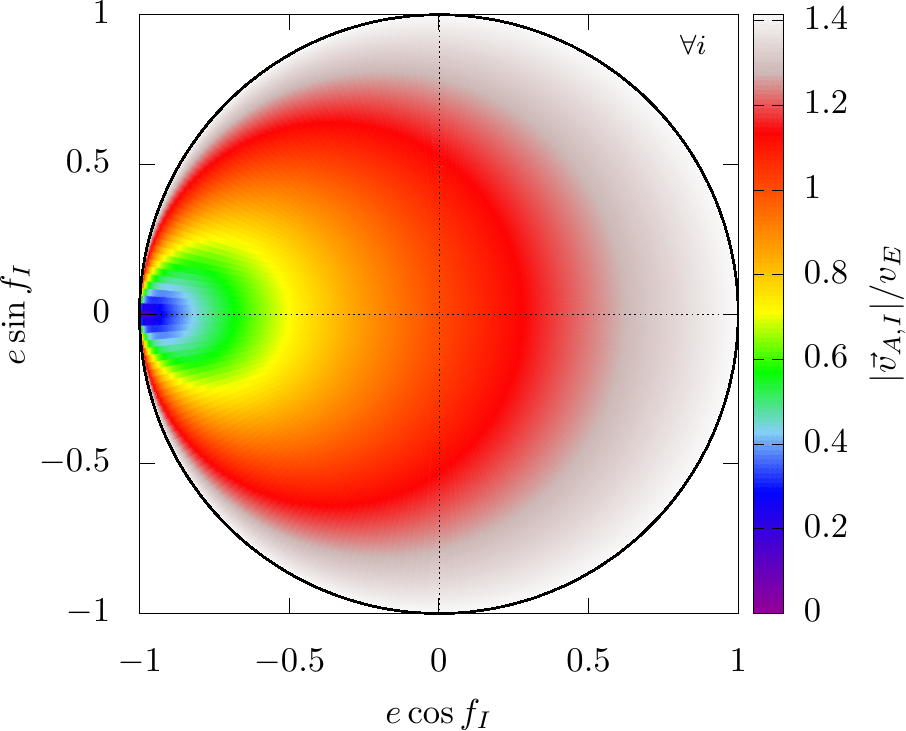}

	\end{center}
		\caption{Norm of the PHA heliocentric velocity at the impact location, in units of the velocity on a circular Earth orbit $v_E$.}
		\label{fig:vIovervE}
	\end{figure}

The heliocentric velocity of any PHA at the impact location ($f = f_I$) can indeed be written in terms of $f_I, e,$ and $i$ as follows: $\vec{v}_{A,I} = v_{A,I,x} \ \vec{e}_x + v_{A,I,y} \ \vec{e}_y + v_{A,I,z} \ \vec{e}_z$, with
\be
	\begin{split}
		v_{A,I,x} &=  \sqrt{\frac{\mu_{\rm Sun}}{\rm 1\hspace{2pt}AU}} \frac{e \sin(f_I)}{\sqrt{ 1 + e \cos(f_I)}}\\
		v_{A,I,y} &=  \sqrt{\frac{\mu_{\rm Sun}}{\rm 1\hspace{2pt}AU}} \sqrt{1 + e \cos(f_I)} \cos(i)\\
		v_{A,I,z} &=  \pm\sqrt{\frac{\mu_{\rm Sun}}{\rm 1\hspace{2pt}AU}} \sqrt{1 + e \cos(f_I)} \sin(i)
	\end{split}
\ee
since $\vec{v}_{A,I} = v_{A,r}(f_I) \ \vec{e}_x + v_{A,f}(f_I) \ (\cos i\ \vec{e}_y	\pm \sin i\ \vec{e}_z)$.
	Its norm, as a function of $f_I$ and $e$:
	\be
		\begin{split}
		|\vec{v}_{A,I}|
				= \sqrt{ \frac{\mu_{\rm Sun}}{\rm 1\hspace{2pt}AU} }  \sqrt{\frac{1 + 2e\cos f_I + e^2}{1 + e \cos f_I} }
				= \sqrt{ \frac{\mu_{\rm Sun}}{\rm 1\hspace{2pt}AU}} \sqrt{2 -  \frac{1 - e^2}{1 + e \cos f_I} }.
		\end{split}
	\ee
	is shown in Fig.~\ref{fig:vIovervE}.
	The norm of course does not depend on the inclination of the asteroid orbit; as well-known, it actually only depends on the orbital period of the asteroid:
	\be
		|\vec{v}_{A,I}|
				= \sqrt{ \frac{\mu_{\rm Sun}}{\rm 1\hspace{2pt}AU}} \sqrt{2 - \frac{1}{ {P_{\rm yr}}^{\frac{2}{3}} } }.
	\ee

\subsubsection*{At the opposite node}

	\begin{figure}[h!!]
	\begin{center}
		\includegraphics[width = .6\textwidth]{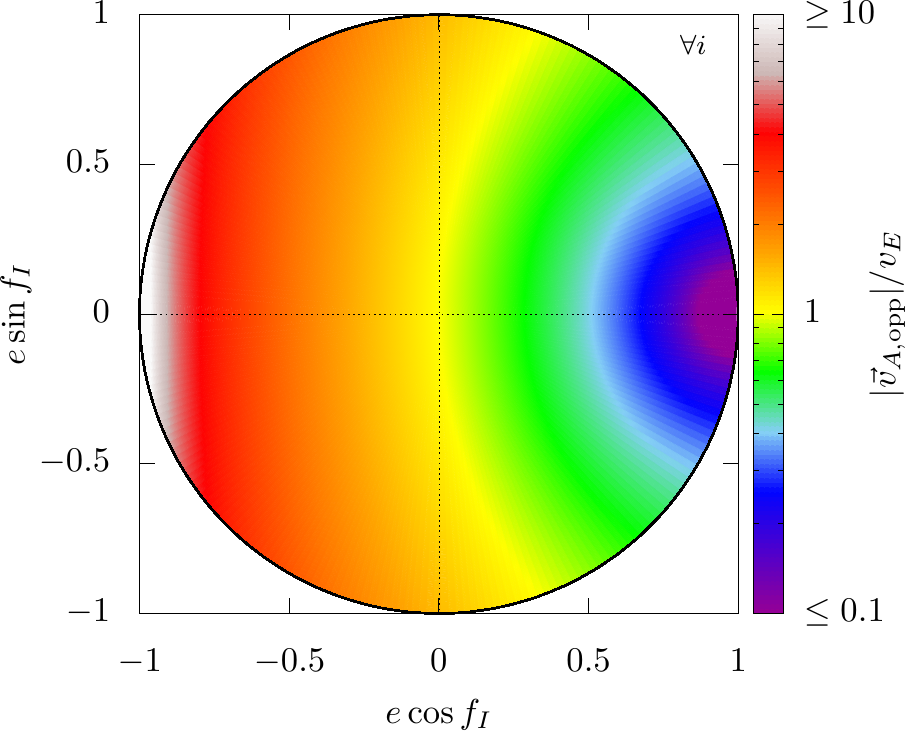}

	\end{center}
		\caption{Same as Fig.~\ref{fig:vIovervE}, but for the velocity at the opposite node. Notice that the scale in this plot is logarithmic.}
		\label{fig:voppovervE}
	\end{figure}

	Similarly, the heliocentric velocity at the opposing node $\vec{r}_{\rm opp} = \vec{r}(\pi + f_I \mod 2\pi)$ reads: $\vec{v}_{A,{\rm opp}} = v_{A,{\rm opp},x} \ \vec{e}_x + v_{A,{\rm opp},y} \ \vec{e}_y + v_{A,{\rm opp},z} \ \vec{e}_z$, with
\be
	\begin{split}
		v_{A,{\rm opp},x} &=  \sqrt{\frac{\mu_{\rm Sun}}{\rm 1\hspace{2pt}AU}} \frac{e \sin(f_I)}{\sqrt{ 1 + e \cos(f_I)}} = v_{A,I,x}\\
		v_{A,{\rm opp},y} &= -\sqrt{\frac{\mu_{\rm Sun}}{\rm 1\hspace{2pt}AU}} \frac{1 - e \cos(f_I)}{ \sqrt{ 1 + e \cos(f_I) } } \cos(i)\\
		v_{A,{\rm opp},z} &= \mp\sqrt{\frac{\mu_{\rm Sun}}{\rm 1\hspace{2pt}AU}} \frac{1 - e \cos(f_I)}{ \sqrt{ 1 + e \cos(f_I) } } \sin(i)
	\end{split}
\ee
since $\vec{v}_{A,{\rm opp}} = v_{A,r}(\pi+f_I) \ (-\vec{e}_x) + v_{A,f}(\pi+f_I) \ (-\cos i\ \vec{e}_y	\mp \sin i\ \vec{e}_z)$.
	Figure~\ref{fig:voppovervE} shows the norm of the heliocentric velocity at the opposite node:
	\be
		\begin{split}
		|\vec{v}_{A,{\rm opp}}|
				&= \sqrt{ \frac{\mu_{\rm Sun}}{\rm 1\hspace{2pt}AU} }  \sqrt{\frac{1 - 2e\cos f_I + e^2}{1 + e \cos f_I} },
		\end{split}
	\ee
	which one could
	also write
	\be
		\begin{split}
		|\vec{v}_{A,{\rm opp}}|
						 &= \sqrt{ \frac{\mu_{\rm Sun}}{|\vec{r}_{\rm opp}|} }  \sqrt{2  - \frac{1 - e^2}{1 - e \cos f_I} }
						 = \sqrt{ \frac{\mu_{\rm Sun}}{|\vec{r}_{\rm opp}|} }  \sqrt{2  - \frac{1}{ {P_{\rm yr}}^{\frac{2}{3}} } \frac{|\vec{r}_{\rm opp}|}{|\vec{r}_I|}    }.
		\end{split}
	\ee
	Again, it is important to keep in mind that, while $\vec{r}_I$ is the same for all PHAs, $\vec{r}_{\rm opp}$ is not: the way it depends on $f_I$ and $e$ is given in Eq.~\eqref{eq:ropp}.

	\paragraph*{Remarks:}

	\begin{itemize}
	\item At both nodes the projection of the heliocentric velocity along the line of nodes ($x$-direction) is actually the same. For a fixed $e$, it is at most $e\ v_E$, with $v_E = \sqrt{\mu_{\rm Sun}\over{\rm 1\hspace{2pt}AU}}$.

	\item At the impact location, the norm $|\vec{v}_{A,I}|$ and the non-radial part of the velocity $v_{A,I,f}$ take values between 0 and the Solar-System escape velocity at 1~AU from the Sun $\sqrt{2}v_E$, as expected for elliptical orbits; for a given $e$, the maximum and minimum correspond to $f_I = 0$ or $\pi$, at which $v_{A,I,x}$ vanishes. On the other hand, remember that the equivalent velocities at the opposing node depend on the scale given by $|\vec{r}_{\rm opp}|$ itself, which changes from one PHA to another: {\it e.g.} $|\vec{v}_{A,{\rm opp}}| < \sqrt{2}\sqrt{\frac{\mu_{\rm Sun}}{|\vec{r}_{\rm opp}(f_I,e)|}}$.
	\end{itemize}

\subsubsection{Flight-path angle}

For completeness, the asteroid flight-path angle at both nodes for any PHA orbit is easily derived.
At $\vec{r}_I$:
\be
	\gamma_I = \atan\big( \frac{e \sin f_I}{1 + e \cos f_I} \big).
\ee
At $\vec{r}_{\rm opp}$:
\be
	\gamma_{\rm opp} = \atan\big( \frac{-e \sin f_I}{1 - e \cos f_I} \big).
\ee
Note that the flight-path angles corresponding to both locations are equal to zero if the impact point corresponds to either the perihelion or the aphelion of the PHA orbit, as they should be.

While $f_I \in {[}0,2\pi{[}$, the flight-path angle goes from $-\frac{\pi}{2}$ to $\frac{\pi}{2}$. Note that, in the limit $e \rightarrow 1$,
we actually have:
\be
	\begin{split}
	\tan \gamma_I \rightarrow \tan\frac{f_I}{2}, \qquad \textrm{namely, if } f_I < \pi, \  \gamma_I &\rightarrow \frac{f_I}{2}\\
														\textrm{while if } f_I > \pi, \ \gamma_I &\rightarrow \frac{f_I - 2\pi}{2}
	\end{split}
\ee
and
\be
	\tan\gamma_{\rm opp} \rightarrow \tan\frac{f_I - \pi}{2}, \qquad {\rm namely} \ \gamma_{\rm opp} \rightarrow \frac{f_I - \pi}{2}.
\ee

\subsection{Loci in the polar plot} \label{sec:polarplot_loci_aph_peri}

\begin{figure}[h!!]
		\begin{center}
			\includegraphics[width = .6\textwidth]{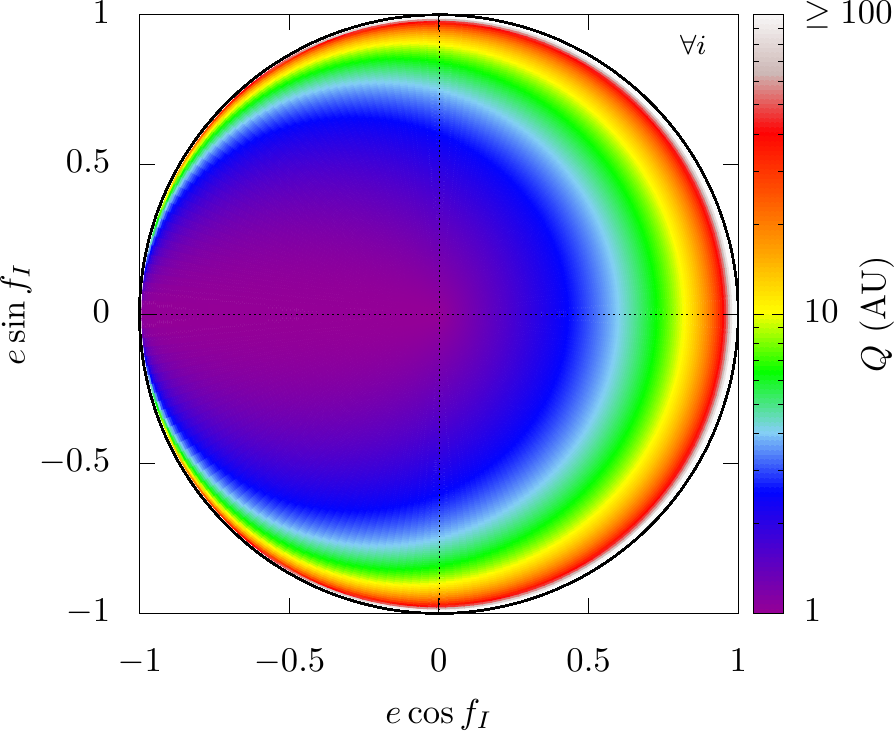}
		\end{center}
		\caption{Aphelion distance of any strictly impacting PHA (notice that the scale in this plot is logarithmic). As expected, orbits that best correspond to the Atiras limit ($Q \rightarrow 1$~AU) lie on the negative abscissa: $f_I = \pi$ (in such a limit, $\vec{r}_I$ would indeed necessarily be the aphelion).}
		\label{fig:aphelion_distance_POLAR}
	\end{figure}

\begin{figure}[h!!]
		\begin{center}
			\includegraphics[width = .6\textwidth]{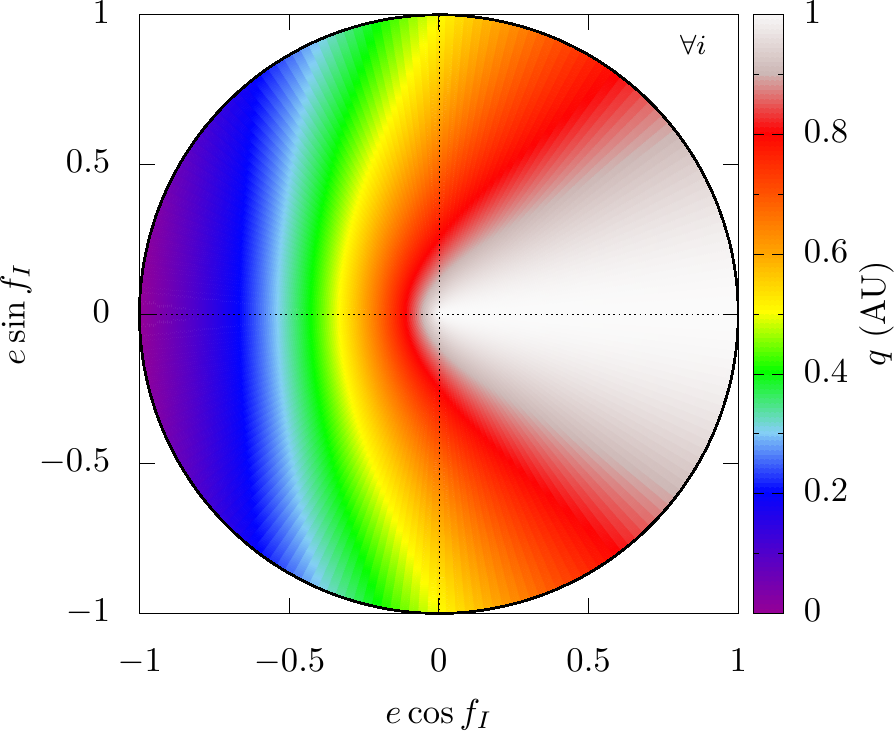}
		\end{center}
		\caption{Perihelion distance of any strictly impacting PHA. As expected, orbits that best correspond to the Amor limit ($q \rightarrow 1$~AU) lie on the positive abscissa: $f_I = 0$ (this time, $\vec{r}_I$ would be the perihelion).}
		\label{fig:perihelion_distance_POLAR}
	\end{figure}

For completeness, as done for the semi-major axis in Sec.~\ref{sec:polar}, we turn to the aphelion and perihelion on the polar plot; see Figs.~\ref{fig:aphelion_distance_POLAR} and~\ref{fig:perihelion_distance_POLAR}.

As a matter of fact, as we did for the Aten--Apollo boundary and the period (resp.\@ the semi-major axis), we can provide an analytical expression for the locus of orbits corresponding to a specific value of the perihelion $q$ or the aphelion $Q$.

\subsection{Aphelion}

Starting from the expression of the aphelion distance:
\be
	Q_{\rm AU} \equiv \left( \frac{Q}{1~{\rm AU}} \right) = \left( \frac{a}{1~{\rm AU}} \right) (1 + e) = \frac{1 + e \cos f_I}{1 - e},
\ee
one can write 
\be
	\left( {Q_{\rm AU}}^2 -1  \right) x^2 + { Q_{\rm AU} }^2 y^2 + 2 \left( Q_{\rm AU} - 1 \right) x - {\left( Q_{\rm AU} - 1 \right)}^2 = 0,
\ee
where $x \equiv e \cos f_I$ and $y \equiv e \sin f_I$.
In the polar plot, this actually corresponds to an ellipse of centre, semi-major axis, and semi-minor axis, respectively given by
\be
	\mathscr{C}_Q = \left( x_0 = -\frac{1}{ Q_{\rm AU} +1},  \ y_0 = 0 \right), \quad \mathscr{A}_Q = \frac{Q_{\rm AU}}{Q_{\rm AU} + 1}, \quad\textrm{and}\quad\mathscr{B}_Q = \sqrt{\frac{Q_{\rm AU}-1}{Q_{\rm AU}+1}}.
	\label{eq:param_locus_Q}
\ee
Note that the eccentricity of any such locus in the polar plot is in fact given by $\mathscr{E}_Q = 1 / Q_{\rm AU}$.

As parametric functions,
all the PHA orbits that share a given aphelion distance $Q_{\rm AU}$ are actually those that correspond to points in the polar plot which satisfy
\be
	\left\{
        \begin{array}{*2{>{\displaystyle}l}p{5cm}}
        	x \equiv e\cos f_I = -\frac{1}{ Q_{\rm AU} +1} + \frac{Q_{\rm AU}}{Q_{\rm AU} + 1} \cos u\\
        	y \equiv e\sin f_I = \sqrt{\frac{Q_{\rm AU}-1}{Q_{\rm AU}+1}} \sin u
        \end{array} \right.
	\qquad \textrm{with } u \in [0,2\pi[.
	\label{eq:aphelion_parametricfcts_POLAR}
\ee

\subsection{Perihelion}

Similarly, let us now provide the parametric equation for the loci of orbits on the polar plot that correspond to PHAs characterised by the same perihelion distance. Starting from
\be
	q_{\rm AU} \equiv \left( \frac{q}{1~{\rm AU}} \right) = \left( \frac{a}{1~{\rm AU}} \right) (1 - e) = \frac{1 + e \cos f_I}{1 + e},
\ee
one finds that each such locus actually corresponds to a branch of an hyperbola:
\be
	\frac{ {(x - x_0)}^2 }{ { \mathscr{A}_q }^2 } - \frac{y^2}{ { \mathscr{B}_q }^2 } = 1,
\ee
where
\be
	x_0 = -\frac{1}{ 1 + q_{\rm AU} }, \quad \mathscr{A}_q = \frac{q_{\rm AU}}{ 1 + q_{\rm AU} }, \quad\textrm{and}\quad\mathscr{B}_q = \sqrt{\frac{1 - q_{\rm AU}}{1 + q_{\rm AU}}};
\ee
notice the close similarities with what we found for the aphelion, in Eq.~\eqref{eq:param_locus_Q}. Also, the eccentricity of any such locus in the polar plot is in fact given by $\mathscr{E}_q = 1 / q_{\rm AU}$.

Again this can be plotted as a parametric function. All the PHA orbits that share a given perihelion distance $q_{\rm AU}$ are actually those that correspond to points in the polar plot\hspace{1pt}---\hspace{1pt}limited to the disk\hspace{1pt}---\hspace{1pt}which satisfy
\be
	\left\{
        \begin{array}{*2{>{\displaystyle}l}p{5cm}}
        	x \equiv e\cos f_I = -\frac{1}{ 1 + q_{\rm AU} } + \frac{q_{\rm AU}}{1 + q_{\rm AU}} \frac{1}{\cos u}\\
        	y \equiv e\sin f_I = \sqrt{\frac{1 - q_{\rm AU}}{1 + q_{\rm AU}}} \tan u
        \end{array} \right.
	\qquad \textrm{with } u \in ]-\frac{\pi}{2},\frac{\pi}{2}[.
	\label{eq:perihelion_parametricfcts_POLAR}
\ee

\subsection{Heliocentric distance to the opposite-node}

All the PHA orbits that share the same $|\vec{r}_{\rm opp}|$ are simply those that correspond to points in the polar plot which satisfy
\be
	\left\{
        \begin{array}{*2{>{\displaystyle}l}p{5cm}}
        	x \equiv e\cos f_I = \frac{|\vec{r}_{\rm opp}| - |\vec{r}_I|}{|\vec{r}_{\rm opp}| + |\vec{r}_I|}\\
        	y \equiv e\sin f_I = \sqrt{1 - x^2} \cos u = \sqrt{1 - {\left[\frac{|\vec{r}_{\rm opp}| - |\vec{r}_I|}{|\vec{r}_{\rm opp}| + |\vec{r}_I|}\right]}^2} \cos u
        \end{array} \right.
	\quad \textrm{with } u \in ]0,2 \pi[.
	\label{eq:ropp_parametricfcts_POLAR}
\ee

\section{Feasibility conditions for a ballistic transfer to $\vec{r}_{\rm opp}$}\label{sec:ropp_and_feasibility}

\begin{figure}
	\begin{center}
		\includegraphics[width = .49\textwidth]{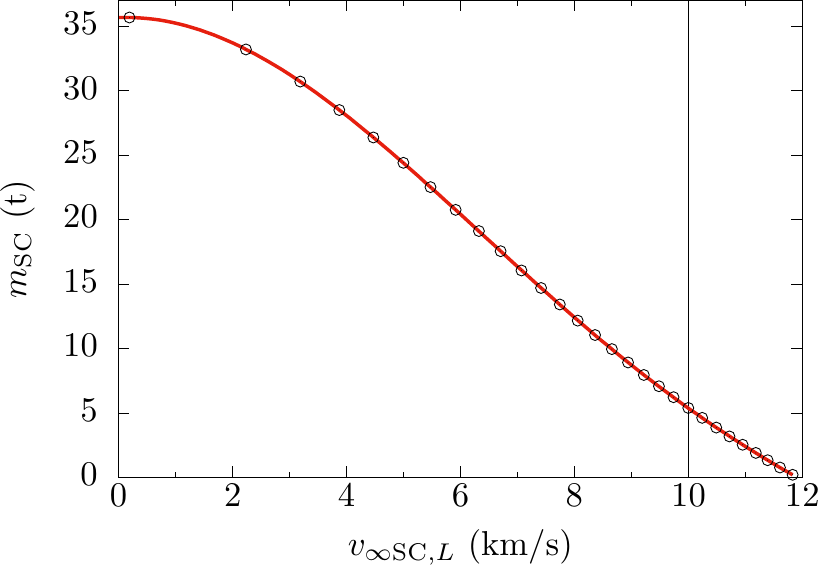}
	\end{center}
		\caption{Launcher performance for the SLS Block-1B 8.4m Fairing EUS~\cite{SLS:2014}.}
		\label{fig:SLSperf}
	\end{figure}

	Knowing what the achievable spacecraft orbits are is important to be able to identify and understand where problems could arise when dealing with parts of the PHA-orbit parameter space. We study this in relation to the launcher performance; ${\max}(|\vec{v}_{\infty {\rm SC},L}|)$ is the largest hyperbolic excess velocity at launch from the Earth system that we are willing to consider for a given launcher (\textit{e.g.\@} about 10 km/s for the SLS Block-1B 8.4m Fairing EUS, see Fig.~\ref{fig:SLSperf}), considering the kinetic-impactor mass trade-off which that value implies.\footnote{We chose the same value as Ref.~\cite{FB:2015}, simply to facilitate the comparison of results with that paper.}

	To determine the largest aphelion and smallest perihelion that could be reached with a given launcher, all needs to be done is to consider the first leg of a Hohmann transfer, with an impulsive $|\Delta \vec{v}|$ equal to ${\max}(|\vec{v}_{\infty {\rm SC},L}|)$ that is directed either along or opposite to the heliocentric Earth velocity at departure. 
	Note that, for clarity, the suffix $\cdot^{\rm (SC)}$ is used when considering properties of the spacecraft orbit. 

Largest ballistically attainable aphelion in the ecliptic plane:
\be
	{\rm max}~Q^{\rm (SC)} = \frac{1+e^{\rm (SC)}}{1-e^{\rm (SC)}}~{\rm AU}, \quad {\rm where} \quad
	e^{\rm (SC)} = -1 + \frac{ 1~{\rm AU} {\left( \sqrt{\frac{\mu_{\rm Sun}}{1~{\rm AU}} } + {\max}(|\vec{v}_{\infty {\rm SC},L}|) \right)}^2 }{ \mu_{\rm Sun} }
	\label{eq:largest_reachable_Q_rocket}
\ee
For instance, for ${\max}(|\vec{v}_{\infty {\rm SC},L}|) = 10$~km/s: ${\rm max}~Q^{\rm (SC)} = 8.268$~AU, $q^{\rm (SC)} = 1$~AU, $e^{\rm (SC)} = 0.784$, $a^{\rm (SC)} = 4.634$~AU, $P^{\rm (SC)} = 9.976$~yr.

Smallest ballistically attainable perihelion in the ecliptic plane:
\be
	{\rm min}~q^{\rm (SC)} = \frac{1-e^{\rm (SC)}}{1+e^{\rm (SC)}}~{\rm AU}, \quad {\rm where} \quad
	e^{\rm (SC)} = 1 - \frac{ 1~{\rm AU} {\left( \sqrt{\frac{\mu_{\rm Sun}}{1~{\rm AU}} } - {\max}(|\vec{v}_{\infty {\rm SC},L}|) \right)}^2 }{ \mu_{\rm Sun} }
	\label{eq:smallest_reachable_q_rocket}
\ee
For instance, for ${\max}(|\vec{v}_{\infty {\rm SC},L}|) = 10$~km/s: ${\rm min}~q^{\rm (SC)} = 0.283$~AU, $Q^{\rm (SC)} = 1$~AU, $e^{\rm (SC)} = 0.559$, $a^{\rm (SC)} = 0.642$~AU, $P^{\rm (SC)} = 0.514$~yr.

	\begin{figure}[h]
	\begin{center}

		\includegraphics[width = .49\textwidth]{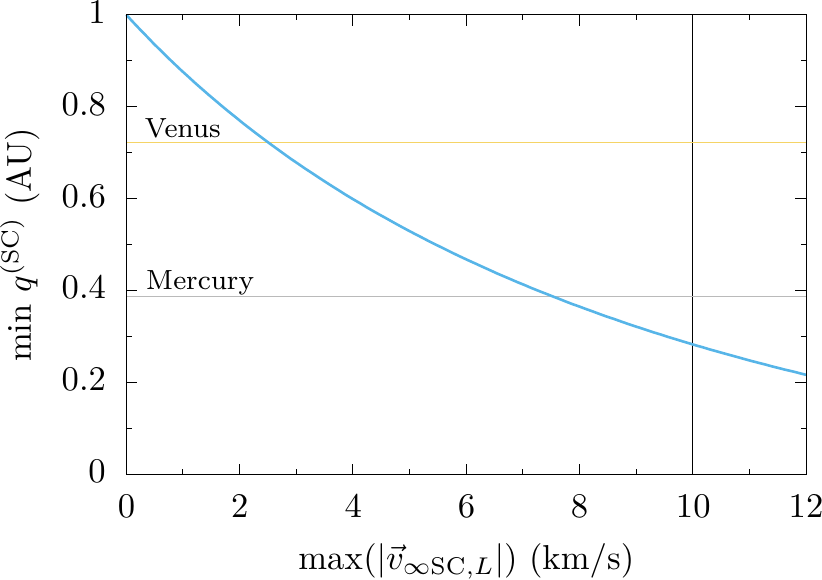}
		\includegraphics[width = .49\textwidth]{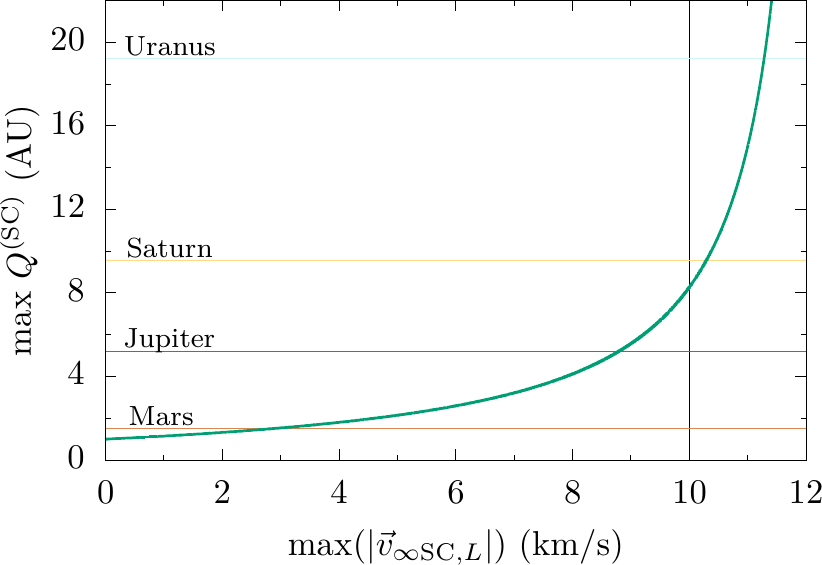}

	\end{center}
		\caption{Smallest (\emph{left}) and largest (\emph{right}) attainable radii in the ecliptic plane with a ballistic trajectory. This should be compared to the radius of the PHA orbit at its opposite node (ecliptic-plane crossing)}
		\label{fig:smallest_reachable_q_and_largest_reachable_Q}
	\end{figure}

The results are shown as functions of the maximum hyperbolic excess velocity at launch in Fig.~\ref{fig:smallest_reachable_q_and_largest_reachable_Q}; to give a sense of scale, we also indicate on these plots the mean semi-major axis for a number of planets.
	This is purely related to the launcher, with no relation with PHAs.

\bigskip

The feasibility conditions for a transfer to the opposite node of any PHA orbit is then very simply determined by comparing 
these achievable locations to the location of the opposite node, given by Eq.~\eqref{eq:ropp}. What is particularly interesting to know is when such a transfer is not feasible.
The condition $|\vec{r}_{\rm opp}| < q_{\rm min}^{\rm (SC)}$ (opposite node so close to the Sun that it cannot be reached) is satisfied whenever
\be
	e^{\rm (PHA)} > \left(\frac{1 - q_{\rm min}^{\rm (SC)}/{\rm AU}}{1 + q_{\rm min}^{\rm (SC)}/{\rm AU}}\right) \frac{1}{-\cos f_I^{\rm (PHA)}};
	\label{eq:ropp_too_small}
\ee
when, on the other hand, the condition $|\vec{r}_{\rm opp}| > Q_{\rm max}^{\rm (SC)}$ (opposite node too far away) holds if
\be
	e^{\rm (PHA)} > \left(\frac{Q_{\rm max}^{\rm (SC)}/{\rm AU} - 1}{Q_{\rm max}^{\rm (SC)}/{\rm AU} + 1}\right) \frac{1}{\cos f_I^{\rm (PHA)}}.
	\label{eq:ropp_too_large}
\ee

In both cases, the first factor in the right-hand side is positive, fixed, and known for a given rocket\hspace{1pt}---\hspace{1pt}see Eqs.~(\ref{eq:largest_reachable_Q_rocket}--\ref{eq:smallest_reachable_q_rocket})\hspace{1pt}---\hspace{1pt}while the second factor, determined by the asteroid orbit, is also always positive, as seen in Fig.~\ref{fig:scenarios_fI}. This result is completely general, without any assumption on whether the impact point is the ascending or descending node, it also holds whether the impact happens as the PHA is heading towards its perihelion or towards its aphelion.

Equations~\eqref{eq:ropp_too_small} and \eqref{eq:ropp_too_large} therefore tell us that a strict $\pi$-transfer is not an option whenever
\be
	e^{\rm (PHA)} > \left(\frac{\mathcal{Q}_{\rm limit}^{\rm (SC)}/{\rm AU} - 1}{1 + \mathcal{Q}_{\rm limit}^{\rm (SC)}/{\rm AU}}\right) \frac{1}{\cos f_I^{\rm (PHA)}}, \qquad \textrm{with $\mathcal{Q}_{\rm limit}^{\rm (SC)}$ being either $q_{\rm min}^{\rm (SC)}$ or $Q_{\rm max}^{\rm (SC)}$.}
\ee
The identification of the two regions (either too close or too far from the Sun) is made in Fig.~\ref{fig:3dpolar_ropp_e_fI} in the specific case ${\max}(|\vec{v}_{\infty {\rm SC},L}|) = 10$~km/s.

	\bigskip

\begin{figure}[h!!]
	\begin{center}
		\includegraphics[width = .6\textwidth]{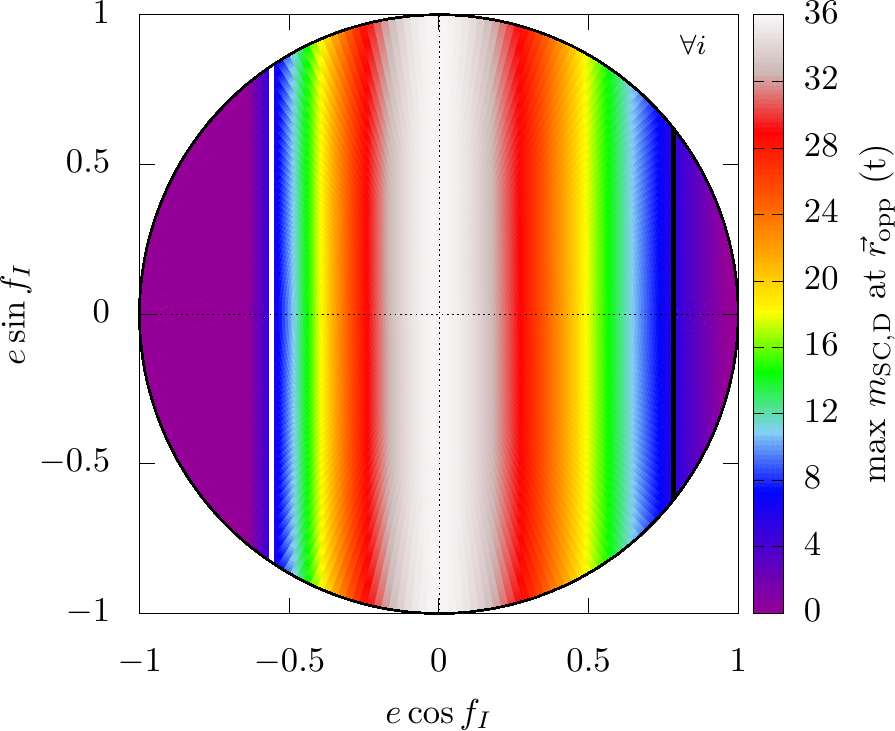}

	\end{center}
		\caption{Maximum kinetic impactor mass $m_{{\rm SC},D}$ in tons that could be brought ballistically
		to the corresponding opposite node of any PHA,
		when considering the performance of the SLS Block-1B 8.4m Fairing EUS~\cite{SLS:2014}, shown in Fig.~\ref{fig:SLSperf}.}
		\label{fig:max_mK_pitransf}
	\end{figure}

	Note that, assuming an explicit performance, it is moreover trivial to determine the maximum kinetic impactor mass that could be sent to $\vec{r}_{\rm opp}$ for any possible asteroid; see {\it e.g.\@} Fig.~\ref{fig:max_mK_pitransf}.

\small

\bibliographystyle{mystyle}
\bibliography{alexbib}

\end{document}